%% file: main.tex
\newif\ifarxiv
\arxivtrue

\ifarxiv
\documentclass[sigconf, nonacm]{acmart}
\else
\documentclass[sigconf, anonymous, review]{acmart}
\fi

\usepackage{preamble}

\setcopyright{none}
\renewcommand\footnotetextcopyrightpermission[1]{}
\acmConference[WWW '27]{The ACM Web Conference 2027}{2027}{}
\acmYear{2027}
\acmDOI{}
\acmISBN{}

\begin{document}

\title[\method{}: Structure-Completed Evidence Units for Web RAG]{\method{}: Structure-Completed Evidence Units for Web Retrieval-Augmented Generation}

\ifarxiv
\author{Peichun Hua}
\affiliation{%
  \institution{The Chinese University of Hong Kong, Shenzhen}
  \city{Shenzhen}
  \country{China}}
\email{peichunhua@link.cuhk.edu.cn}

\author{He Sun}
\affiliation{%
  \institution{The University of Hong Kong}
  \city{Hong Kong}
  \country{China}}
\email{hesun@hku.hk}

\author{Zirui Zhuang}
\affiliation{%
  \institution{Beijing University of Posts and Telecommunications}
  \city{Beijing}
  \country{China}}
\email{zhuangzirui@bupt.edu.cn}

\author{Qi Qi}
\affiliation{%
  \institution{Beijing University of Posts and Telecommunications}
  \city{Beijing}
  \country{China}}
\email{qiqi8266@bupt.edu.cn}

\author{Yiming Qiu}
\affiliation{%
  \institution{The University of Hong Kong}
  \city{Hong Kong}
  \country{China}}
\email{yimingq@hku.hk}

\author{Shinan Liu}
\affiliation{%
  \institution{The University of Hong Kong}
  \city{Hong Kong}
  \country{China}}
\email{shinan6@hku.hk}

\author{Yunming Xiao}
\authornote{Corresponding author.}
\affiliation{%
  \institution{The Chinese University of Hong Kong, Shenzhen}
  \city{Shenzhen}
  \country{China}}
\email{yunmingxiao@cuhk.edu.cn}

\renewcommand{\shortauthors}{Hua et al.}
\fi

\input{sections/abstract}

\ccsdesc[500]{Information systems~Retrieval models and ranking}
\ccsdesc[500]{Information systems~Web searching and information discovery}
\ccsdesc[300]{Information systems~Question answering}

\keywords{Retrieval-Augmented Generation, Content Extraction, Web Search, HTML, Structured Evidence, Agentic Search}

\maketitle

\input{sections/intro}
\input{sections/motivation}
\input{sections/method}
\input{sections/experiments}
\input{sections/related}
\input{sections/conclusion}

\bibliographystyle{ACM-Reference-Format}
\bibliography{references}

\appendix
\input{sections/appendix}

\end{document}

%% file: sections/abstract.tex
\begin{abstract}
Web retrieval-augmented generation (RAG) splits a page into fixed-size text chunks or, in recent visual pipelines, screenshot tiles, and the retrieved item serves at once as the unit of indexing, ranking, and reading. We trace answer evidence from source HTML to reader input over the full English Wikipedia and find that this coupling loses evidence in text and screenshot pipelines alike: evidence present in the source page often fails to reach the reader intact, particularly structured content such as tables and infoboxes, which chunking separates from the headers and context that give it meaning. Stronger extraction, context expansion, and larger reader budgets leave the loss in place. We present \method{}, a training-free evidence delivery layer that decouples what a pipeline retrieves from what its reader reads. Given the pages behind any ranked list, \method{} rebuilds their HTML into fine-grained \emph{\units{}}, each completed with its page title, section path, and table headers, selects the units relevant to the query, and packs them into the reader context at an unchanged token budget. The index, the retriever, and the reader stay as they are, so \method{} applies to text, screenshot, and agentic pipelines. Across three question answering benchmarks, \method{} improves answer accuracy by 5 to 6 points on text and screenshot pipelines, consistently across readers, retrievers, and reader budgets, and outperforms strong HTML-aware and structure-aware RAG baselines under matched budgets. The gains carry over to multi-step agentic search. Our code is publicly available at \url{https://github.com/sarendis56/Tessera}.
\end{abstract}

%% file: sections/intro.tex
\section{Introduction}
\label{sec:intro}

Retrieval-augmented generation (RAG) grounds a language model in an external corpus~\cite{guu2020realm,lewis2020retrieval,karpukhin2020dense,izacard2021leveraging}, and the Web is the corpus that matters most in practice: search assistants, deep-research agents, and browsing models all answer from retrieved webpages~\cite{openai-agent,nakano2021webgpt,vu2023freshllms,jin2025searchr1,song2025r1searcher,liu2026agenticr,wu2025webwalker,hua2026semantics}. A webpage is a structured document. Its facts sit in infoboxes, tables, lists, and captions as often as in prose, and each fact takes its meaning from the surrounding structure: a number in a scorecard means nothing without its column header, its row label, and the section that names the match. Carrying this structure from page source to reader is a Web-specific challenge for RAG and agentic search, which prior work largely overlooks by retrieving from clean passage corpora that strip tables, infoboxes, and lists~\cite{jin2025flashrag,jin2025searchr1,liu2026agenticr}.

\input{figures/teaser}

The standard web RAG pipeline discards this structure early: an extractor such as Trafilatura~\cite{barbaresi2021trafilatura} flattens each page to text, a chunker cuts the text into fixed windows, and the reader receives the top few chunks after retrieval and reranking. HTML-centric methods keep pruned markup in the reader's context~\cite{tan2025htmlrag}. Vision-centric methods skip parsing and index rendered pages as screenshot tiles~\cite{ma2024dse,faysse2025colpali,yu2025visrag}; PixelRAG scales this design to the full English Wikipedia and reports that screenshots beat text across six benchmarks~\cite{wang2026pixelrag}.

We start from the text and screenshot pipelines released with PixelRAG and ask where each loses its answers. Once both rerank their candidates, the two modalities perform on par (\S\ref{sec:gap}), and they share a ceiling that a four times larger reader budget lifts by about two points (\S\ref{sec:remedies}). Tracing quoted answer evidence from the source HTML to the reader input (\S\ref{sec:trace}) shows that it disappears the most at two places.
For 9.7\% of questions, a single stored chunk contains the complete evidence, yet it is missing from the top three after ranking; placing that chunk in the third slot lifts accuracy on these questions from 49.1\% to 78.5\%.
For 7.6\% of source-supported questions, the evidence never enters the index intact, and \textit{81\% of these losses are table cells, table headers, or captions} (Figure~\ref{fig:teaser}).

Both losses follow from \textit{a pipeline that fixes, at indexing time, what its reader can see}. Evidence that extraction drops is lost to every later stage, and the chunk size is tuned for retrieval, where 1,024 tokens is a good choice: in our pilot study, 512-token chunks enlarge the index by 73\% and answer fewer questions (\S\ref{sec:budget}), and the page that supports the answer is among the pages of the top ten reranked chunks for 92.7\% of source-supported questions. The pipeline then hands the same chunks to the reader, and \textbf{a chunk sized for retrieval is a poor unit to read from}. The answer usually occupies one sentence or one table row, so only a few chunks fit in the budget and the one that holds the evidence often falls outside them, and a fixed window separates a table row from the header and caption that label it. \textbf{The pipeline uses a single object, the chunk or the tile, as its unit of indexing, ranking, and reading}: it finds the right page and delivers the wrong part.

We present \method{}\footnote{A \emph{tessera} is a single small tile of a mosaic.}, a training-free, query-time layer that decouples delivery from retrieval on top of any pipeline. It treats the ranked list as a pointer to \emph{anchor pages}, parses their HTML into small \emph{\units{}} such as a table row, a list item, or a paragraph, and completes each unit with the structure that gives it meaning: the page title, the section path, and the column headers resolved through merged cells. A cross-encoder selects the units relevant to the question, and a packer writes them into the token budget of one evidence item, so the reader budget, the index, the retriever, and the reader stay unchanged. Our contributions are as follows.

\begin{itemize}[leftmargin=*,topsep=2pt]
\item \textbf{A source-to-reader evidence trace of web RAG.} We locate where text and screenshot pipelines lose answer evidence, price the loss with a controlled oracle intervention, and show that stronger extraction, context expansion, structure-aware reranking, and larger budgets do not naturally recover it (\S\ref{sec:motivation}).
\item \textbf{\method{}, a universal evidence delivery layer.} We introduce structure-completed \units{}, a faithful table-grid construction, and budget-neutral packing; the layer plugs into text, screenshot, and agentic pipelines without retraining (\S\ref{sec:method}). We show that this is a simple and effective add-on to all of traditional RAG, multi-modal index, and agentic RAG (\S\ref{sec:main}).
\item \textbf{Consistent gains across pipelines and settings.} On 7,851 questions from Natural Questions~\cite{kwiatkowski2019natural}, NQ-Tables~\cite{herzig2021nqtables}, and SimpleQA~\cite{wei2024simpleqa}, \method{} adds 5 to 6 points to text and screenshot pipelines, and up to 5.2 points inside a multi-step search agent on FRAMES~\cite{krishna2025frames} and MuSiQue~\cite{trivedi2022musique}.
\ifarxiv
Under a matched budget, it outperforms all fourteen alternatives we test: four query-expansion methods~\cite{abduljaleel2004rm3,lassance2024spladev3,gao2023hyde,wang2023query2doc}, two document-expansion methods~\cite{nogueira2019doctttttquery,gospodinov2023doc2query}, late chunking~\cite{gunther2024late}, small-to-big retrieval~\cite{liu2022llamaindex}, RAPTOR~\cite{sarthi2024raptor}, contextual retrieval~\cite{anthropic2024contextual}, two variants of HtmlRAG~\cite{tan2025htmlrag}, and two structure-aware chunkers~\cite{unstructured,auer2024docling}.
\fi
The gain persists across readers, retrievers, chunk sizes, and reader budgets, and 3K tokens with \method{} beat 12K tokens of chunks or tiles (\S\ref{sec:experiments}).
\end{itemize}

%% file: figures/teaser.tex
\begin{figure}[t]
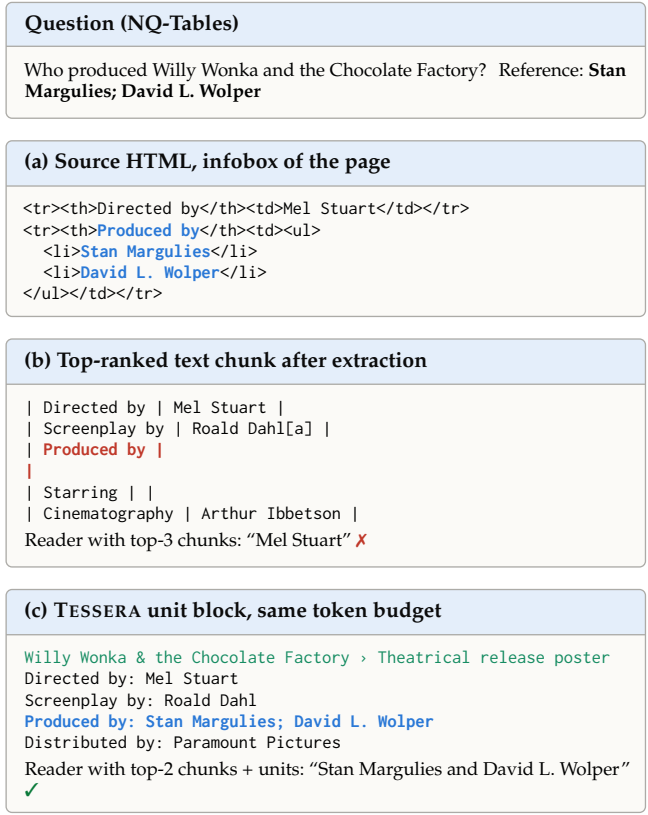

\centering
\begin{examplebox}{Question (NQ-Tables)}
\rdr{Who produced Willy Wonka and the Chocolate Factory?\hfill Reference: \textbf{Stan Margulies; David L. Wolper}}
\end{examplebox}
\vspace{-3pt}
\begin{examplebox}{(a) Source HTML, infobox of the page}
<tr><th>Directed by</th><td>Mel Stuart</td></tr>\\
<tr><th>\hl{Produced by}</th><td><ul>\\
\hspace*{1em}<li>\hl{Stan Margulies}</li>\\
\hspace*{1em}<li>\hl{David L. Wolper}</li>\\
</ul></td></tr>
\end{examplebox}
\vspace{-3pt}
\begin{examplebox}{(b) Top-ranked text chunk after extraction}
| Directed by | Mel Stuart |\\
| Screenplay by | Roald Dahl[a] |\\
| \hlbad{Produced by |}\\
\hlbad{|}\\
| Starring | |\\
| Cinematography | Arthur Ibbetson |\\[2pt]
\rdr{Reader with top-3 chunks: ``Mel Stuart'' \xmark}
\end{examplebox}
\vspace{-3pt}
\begin{examplebox}{(c) \method{} unit block, same token budget}
\ctx{Willy Wonka \& the Chocolate Factory › Theatrical release poster}\\
Directed by: Mel Stuart\\
Screenplay by: Roald Dahl\\
\hl{Produced by: Stan Margulies; David L. Wolper}\\
Distributed by: Paramount Pictures\\[2pt]
\rdr{Reader with top-2 chunks + units: ``Stan Margulies and David L. Wolper'' \cmark}
\end{examplebox}
\caption{Evidence lost between a webpage and the reader: the extractor drops a list-valued infobox cell, and \method{} delivers the completed row in place of the third chunk.}
\label{fig:teaser}
\end{figure}

%% file: sections/motivation.tex
\section{Where Does Web Evidence Go Missing?}
\label{sec:motivation}

We first show that text and screenshot pipelines converge once both rank their candidates well (\S\ref{sec:gap}). We then trace answer evidence through the text pipeline and quantify what the reader never sees (\S\ref{sec:trace}), test the obvious remedies, and derive the design principles of \method{} (\S\ref{sec:principles}). All experiments use the Qwen3.5-4B reader and August 2025 English Wikipedia snapshot, consistent with our main experiment setup in \S\ref{sec:setup}.

\subsection{The Modality Gap Is a Ranking Gap}
\label{sec:gap}

PixelRAG compares a screenshot retriever against a text baseline that sends the top three results of a single dense retriever to the reader~\cite{wang2026pixelrag}. Table~\ref{tab:gap} reproduces this comparison on PixelRAG's released indexes and then gives each modality a reranker over its top 50 candidates: BGE-reranker-v2-m3~\cite{chen2024bgem3} for text chunks and Qwen3-VL-Reranker~\cite{li2026qwenvlretrieval} for screenshot tiles.

\input{tables/gap}

Without reranking, screenshots lead text by 3.9 to 5.2 accuracy points, in line with the PixelRAG report. Reranking adds 8.9 to 11.8 points to the text pipeline and 5.2 to 9.5 points to the screenshot pipeline, and the two modalities meet: the difference is $-0.4$ points on \nq, $+1.4$ on \nqt, and $+2.6$ on \sqa. A text pipeline with our own index, which fuses dense and BM25 candidates~\cite{robertson2009bm25,cormack2009rrf} before the same reranker, is the most accurate on every dataset.

The comparison also exposes a shared ceiling. The best pipeline of either modality retrieves the gold article in its top three for 61.8\% of \nqt questions and answers 48.4\% of them. Better ranking has brought both modalities to the same place, and the remaining errors call for a different explanation.

\subsection{Tracing Evidence from Source to Reader}
\label{sec:trace}

\para{Protocol.} We trace the hybrid text pipeline of Table~\ref{tab:gap} on 7,378 questions.\footnote{The trace covers the 7,851 evaluation questions minus 473 whose source pages we used while developing the protocol.} For each question, a grader model (GPT-5.6 Luna) reads the Wikipedia page that the dataset cites and quotes the fragments that support the reference answer. We then locate every quoted fragment at each stage by string matching: in the extracted text, in the stored 1,024-token chunks, in the top-100 retrieved candidates, in the 50 candidates the reranker scores, and in the three chunks the reader receives. The source page supports the reference answer with valid quotes for 5,883 questions, and the trace assigns each of them to the first stage at which its evidence disappears.

\input{figures/trace}

\para{Stored evidence misses the reader.} Figure~\ref{fig:trace} shows the outcome. The quoted evidence reaches the reader for 59.5\% of the source-supported questions. For 16.5\% the evidence sits verbatim in stored chunks and disappears afterwards: the reader's top three omit it for 500 questions, the top-100 candidates omit it for 362, and the reranker's admission pool omits it for 110. The loss is largest on \nq, where it affects one supported question in four.

\para{Oracle intervention.} To measure what this evidence is worth, we select the 715 questions (9.7\% of traced questions) for which one stored chunk contains every quoted fragment and the reader input lacks the evidence. We keep the reader's first two chunks and replace the third with that chunk, so the reader still receives three slots. Table~\ref{tab:oracle} reports the result: accuracy rises from 49.1\% to 78.5\%, with 226 questions gained and 16 lost. Without any oracle information, \method{} already closes more than half of this gap (\S\ref{sec:analysis}). As a control, another chunk from the same page moves accuracy by only 2.1 points, so the answer-bearing passage carries the effect.

\input{tables/oracle}

\para{Extraction drops structured evidence.} For a further 20.3\% of supported questions, the quoted fragments do not appear verbatim in any stored chunk. Tolerant matching, which ignores case, punctuation, whitespace, and word order, shows that most paragraph fragments survive in reformatted form, while structured content does not: extraction loses 16\% of quoted table cells, 40\% of figure captions, and 44\% of table captions, compared to 1\% of paragraph fragments. In total, 449 questions (7.6\% of the supported ones) lose at least one evidence fragment from every stored chunk, and for 362 of them (81\%) the missing fragment is a table cell, a table header, or a caption. No retriever or reranker can recover this evidence because the index does not contain it.

\para{Surviving structure loses its labels.} Table evidence that does survive often arrives stripped of meaning. Among quoted table cells found verbatim in a chunk, 31\% sit in a chunk that lacks the table's leading column headers. Later chunks of a page also lose its title.

\para{Standard remedies fall short.}
\label{sec:remedies}
Each loss suggests a familiar remedy: recall-oriented re-extraction, neighboring chunks around the hits, structure prefixed to the chunks, a four times larger reader budget, or screenshots. Each adds at most 2.2 points with noticeable increased budget or inference-time overhead, against the 29.4 points that the oracle chunk shows to be available (details in Appendix~\ref{app:remedies}).

\subsection{Design Principles}
\label{sec:principles}

The measurements point to four principles.
\textbf{P1: Retrieve pages, deliver units.} The top ten reranked chunks reach the gold page for 92.7\% of supported questions but deliver the evidence for 59.5\%, so the delivery step should reopen the retrieved pages and choose what to read at a finer grain than the index stores.
\textbf{P2: A unit must carry its structure.} A table row is useful only with its headers, and a list item only with the heading above it (\S\ref{sec:ablation}, Appendix~\ref{app:ablation}).
\textbf{P3: Work from the page source at query time.} Extraction loss is invisible downstream of the index, whereas parsing the HTML of a handful of pages per query sees the full page, costs little, and leaves the index untouched, whether it holds text or screenshots.
\textbf{P4: Stay within the reader budget.} One well-chosen slot is worth 29 points on the affected questions and many poorly chosen ones are worth two, so the layer should reallocate the existing budget.

%% file: tables/gap.tex
\begin{table}[t]
\caption{Text and screenshot pipelines on the full Wikipedia: answer accuracy / Recall@3 of the gold article (\%).}
\label{tab:gap}
\centering
\setlength{\tabcolsep}{3.2pt}
\begin{tabular}{@{}lccc@{}}
\toprule
\headrow \textbf{Pipeline} & \textbf{\nq} & \textbf{\nqt} & \textbf{\sqa} \\
\midrule
\grouprow{4}{Released PixelRAG indexes, first-stage retrieval only}
Text, dense & 53.2 / 47.1 & 37.1 / 37.7 & 68.3 / 80.1 \\
Screenshot, tuned & 57.1 / 58.1 & 42.2 / 51.0 & 73.3 / 85.3 \\
\grouprow{4}{Released PixelRAG indexes, reranked top 50}
Text + BGE & 62.7 / 66.5 & 46.0 / 59.2 & 80.1 / 87.5 \\
Screenshot + Qwen3-VL & 62.3 / 62.6 & 47.4 / 57.9 & 82.7 / 89.4 \\
\grouprow{4}{Our index}
Hybrid text + BGE & \best{64.8} / \best{67.8} & \best{48.4} / \best{61.8} & \best{83.6} / \best{90.1} \\
\bottomrule
\end{tabular}
\end{table}

%% file: figures/trace.tex
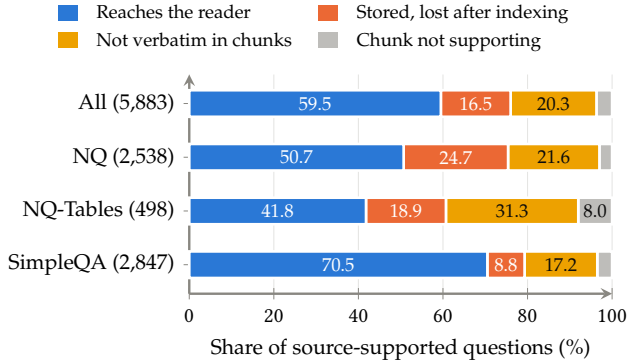
\begin{figure}[t]
\centering
\begin{tikzpicture}
\begin{axis}[
  paperplot, xbar stacked, axis lines*=left, ymajorgrids=false, xmajorgrids,
  scale only axis, width=0.66\columnwidth, height=2.9cm, bar width=10pt,
  xmin=0, xmax=100, xlabel={Share of source-supported questions (\%)},
  ytick={0,1,2,3},
  yticklabels={{\sqa{} (2,847)}, {\nqt{} (498)}, {\nq{} (2,538)}, {All (5,883)}},
  enlarge y limits=0.18,
  legend style={at={(0.30,1.06)}, anchor=south, legend columns=2, column sep=5pt, font=\figfont\footnotesize, /tikz/every even column/.append style={column sep=8pt}},
  legend image code/.code={\draw[#1, draw=none] (0cm,-0.09cm) rectangle (0.26cm,0.11cm);},
  nodes near coords, nodes near coords style={font=\figfont\footnotesize, text=white},
  point meta=explicit symbolic,
]
\addplot[fill=cOurs, draw=white, line width=1pt] coordinates {(70.5,0)[70.5] (41.8,1)[41.8] (50.7,2)[50.7] (59.5,3)[59.5]};
\addplot[fill=cBase, draw=white, line width=1pt] coordinates {(8.8,0)[8.8] (18.9,1)[18.9] (24.7,2)[24.7] (16.5,3)[16.5]};
\addplot[fill=cAltB, draw=white, line width=1pt, nodes near coords style={text=cInk}] coordinates {(17.2,0)[17.2] (31.3,1)[31.3] (21.6,2)[21.6] (20.3,3)[20.3]};
\addplot[fill=cMute!55, draw=white, line width=1pt, nodes near coords style={text=cInk}] coordinates {(3.5,0)[] (8.0,1)[8.0] (3.0,2)[] (3.7,3)[]};
\legend{Reaches the reader, {Stored, lost after indexing}, Not verbatim in chunks, Chunk not supporting}
\end{axis}
\end{tikzpicture}
\caption{Where quoted answer evidence stops in the hybrid text pipeline. ``Chunk not supporting'' means that the quotes sit in a chunk that the grader does not judge supporting.}
\label{fig:trace}
\end{figure}

%% file: tables/oracle.tex
\begin{table}[t]
\caption{Oracle intervention: the third reader slot is replaced by a stored chunk that holds the complete evidence (71 tokens longer on average). Accuracy in \%.}
\label{tab:oracle}
\centering
\setlength{\tabcolsep}{4.5pt}
\begin{tabular}{@{}lrccc@{}}
\toprule
\headrow \textbf{Dataset} & \textbf{$n$} & \textbf{Top-3} & \textbf{\method{}} & \textbf{Oracle} \\
\midrule
\nq & 478 & 51.9 & 61.9 & \best{75.3} \\
\nqt & 62 & 48.4 & 54.8 & \best{71.0} \\
\sqa & 175 & 41.7 & 77.1 & \best{89.7} \\
\midrule
All & 715 & 49.1 & 65.0 & \best{78.5} \\
\bottomrule
\end{tabular}
\end{table}

%% file: sections/method.tex
\section{\method{}}
\label{sec:method}

\input{figures/overview-fig}

\method{} is a delivery layer between a retrieval pipeline and its reader (Figure~\ref{fig:overview}). It consumes the pipeline's ranked list and produces a reader context of the same size in which one evidence item is replaced by a block of structure-completed \units{}. We first state the problem (\S\ref{sec:problem}), then describe the four steps: anchoring (\S\ref{sec:anchor}), unit construction (\S\ref{sec:units}), selection (\S\ref{sec:selection}), and packing (\S\ref{sec:packing}). \S\ref{sec:plug} explains how the layer attaches to text pipelines, screenshot pipelines, and search agents.

\subsection{Problem Setting}
\label{sec:problem}

A retrieval pipeline answers a query $q$ with a ranked list $L=(e_1,\dots,e_n)$ of evidence items. An item is a text chunk or a screenshot tile, and each item $e_i$ comes from a webpage $\pi(e_i)$. The reader receives the top $k$ items, at a cost of $B=\sum_{i\le k}|e_i|$ tokens under the reader's tokenizer or image processor. We call $C=(e_1,\dots,e_k)$ the \emph{coupled} context, because its elements were sized for indexing.

\method{} builds a \emph{decoupled} context
\begin{equation}
C' = (e_1,\dots,e_{k-1},\,U(q,L)), \qquad |U(q,L)| \le |e_k|,
\end{equation}
where $U(q,L)$ is a text block assembled from units of the pages behind $L$. The context keeps the pipeline's best $k-1$ items, which carry continuous prose and the pipeline's ranking signal, and spends the last slot on units. The cost stays at most $B$. The layer is training-free and reads only $q$, $L$, and the HTML of pages $\pi(e_i)$.

\subsection{Anchor Pages}
\label{sec:anchor}

The ranked list is a noisy pointer to passages and a reliable pointer to pages (\S\ref{sec:principles}). \method{} takes the top $m$ items of $L$ and collects their distinct pages in rank order,
\begin{equation}
\mathcal{P}(q) = \mathrm{distinct}\big(\pi(e_1),\dots,\pi(e_m)\big),
\end{equation}
with $m=10$ by default. Several top items often share a page, so $\mathcal{P}(q)$ holds at most $m$ pages. The layer then fetches the HTML of each anchor page from the page store, which is the crawl snapshot that the index was built from. Units are query-independent, so a deployment can cache them per page.

\subsection{Structure-Completed Units}
\label{sec:units}

A unit is the smallest piece of a page that states a fact once its context is attached. Each unit $u$ has a context $c(u)$ (page title, section path, table title, group label), optional labels $h(u)$ (column headers), and a body $b(u)$. Its \emph{completed} text is $c(u)$ followed by $b(u)$, with each row value paired with its label, and is self-contained: a reader, or a reranker, can interpret it without the rest of the page. \method{} walks the DOM of an anchor page in document order after removing scripts, styles, navigation, and citation superscripts, tracking the section headings above the current position. The walk emits two kinds of units, about 150 per Wikipedia page.

\para{Block units.} Each paragraph, list item, definition term or value, and figure caption becomes one unit. Its context $c(u)$ is the path \textit{page title › section › subsection} given by the heading stack.

\para{Row units.} Each row of a table or an infobox becomes one unit. The difficulty is to attach every cell to the right label. Web tables use merged cells, multi-row headers, title rows, and group rows, and the cells of a row in DOM order do not line up with the header cells above them. On the anchor pages of our experiments, pairing cells with the first header row in DOM order assigns a wrong header to 32.5\% of the 2.9 million table rows, or drops one. \method{} therefore rebuilds each table as a \emph{faithful grid} before reading rows from it (Figure~\ref{fig:grid}):
\begin{enumerate}[leftmargin=*,topsep=2pt]
\item \emph{Span expansion.} Every cell is copied into all grid positions that its \texttt{rowspan} and \texttt{colspan} cover, so each grid position holds the cell that a reader of the rendered table would see there.
\item \emph{Title rows.} The caption and the leading rows that span the full width become the table title and join $c(u)$.
\item \emph{Header rows.} The following rows that consist of header cells form $h(u)$. For multi-row headers, the labels of a column are joined top-down, for example \textit{Term of office › Took office}.
\item \emph{Group rows.} A full-width row inside the body labels the rows below it and joins their $c(u)$ until the next group row.
\item \emph{Key--value rows.} In a table without column headers whose rows start with a header cell (an infobox), the first cell is the key.
\item \emph{Nested tables} inherit the title and headers of the table that wraps them. Line breaks and list items inside a cell are kept as separators, preserving multi-valued cells (Figure~\ref{fig:teaser}).
\end{enumerate}

\input{figures/grid}

\subsection{Unit Selection}
\label{sec:selection}

The anchor pages of a question hold about a thousand units, and one slot holds a few dozen. Selection runs in two stages. A lexical prefilter keeps the $N=100$ units whose completed text shares the most content words with the question, which costs no model call. A cross-encoder then scores each remaining pair $(q,\text{completed}(u))$; we reuse the pipeline's own text reranker, BGE-reranker-v2-m3, so the layer adds no model to the deployment. Scoring the completed text matters: the context states which match, season, or entity a row belongs to, and the bare row does not reveal it (compare the two scorecards in Figure~\ref{fig:block}).

\subsection{Organized Budget-Neutral Packing}
\label{sec:packing}

The packer fills the budget $|e_k|$ of the displaced item. Completion serves the scorer, which judges each unit in isolation; the reader sees the selected units together and needs each context only once, so \method{} groups units by context instead of repeating the same path and headers on every row. It visits units in descending score order and admits a unit if its cost fits the remaining budget, where the cost of a unit is its body plus, for the first unit of a group, the group's context line and header line. After selection it renders each group once: the context path, then the column headers, then the member rows as value lines in page order. Groups appear in the order of their best unit. The result reads like an excerpt of the original table (Figure~\ref{fig:block}), and it fits about 14 lines into a slot that holds about 11 units when each one repeats its context.

\input{figures/block}

The packer measures cost with the reader's tokenizer, so the unit block never exceeds the displaced item. In our runs the mean reader prompt with \method{} is within 12 tokens of the baseline prompt.

\subsection{Plugging into Pipelines}
\label{sec:plug}

\para{Text pipelines.} The anchor pages are the pages of the top reranked chunks, and the unit block replaces the last chunk. Because units come from the page HTML, they include content that the pipeline's extractor dropped before indexing.

\para{Screenshot pipelines.} A screenshot index stores each tile with its URL and coordinates. \method{} anchors on the pages of the top tiles and replaces the last tile with the unit block, packed to the reader tokens that the image processor charges for that tile. The block works as text or as a rendered image (Appendix~\ref{app:pixel}).

\para{Search agents.} An agent issues several queries and reads a tool result for each. \method{} applies to every search call independently with the agent's query in the role of $q$: the tool returns the top $k-1$ items and the unit block as the last document. The agent's prompt, tool schema, and stopping rule stay unchanged.

\para{Cost.} Per query, the layer parses the anchor pages, scores $N$ short pairs, and tokenizes the selection. Building the units of a page takes 15\,ms on one CPU core, and scoring takes 0.2 GPU-seconds per question on an A100. The layer adds no index, no offline pass over the corpus, and no reader token budget.

%% file: figures/overview-fig.tex
\begin{figure}[t]
\centering
\includegraphics[width=\columnwidth]{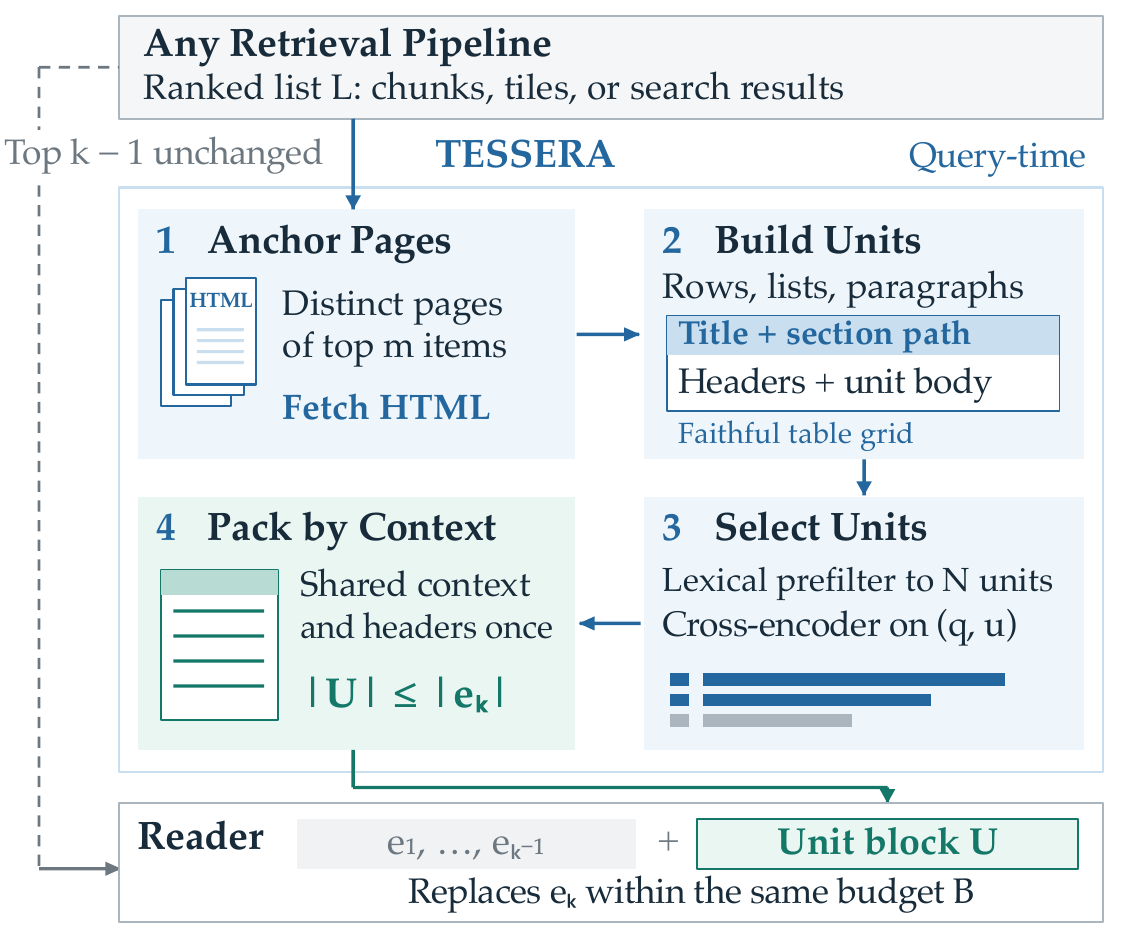}
\caption{Overview of \method{}: retrieved pages are rebuilt into \units{} and packed into the last evidence slot.}
\label{fig:overview}
\end{figure}

%% file: figures/grid.tex
\begin{figure}[t]
\centering
\begin{examplebox}{(a) Rendered table with merged cells (illustrative)}
\centering
{\figfont\footnotesize
\setlength{\tabcolsep}{5pt}\renewcommand{\arraystretch}{1.05}
\begin{tabular}{|l|l|l|l|}
\hline
\multicolumn{4}{|c|}{\textbf{Governors}}\\\hline
\multirow{2}{*}{\textbf{Name}} & \multicolumn{2}{c|}{\textbf{Term of office}} & \multirow{2}{*}{\textbf{Party}}\\\cline{2-3}
 & \textbf{Took office} & \textbf{Left office} & \\\hline
A. Reyes & 1998 & 2004 & \multirow{2}{*}{Liberal}\\\cline{1-3}
B. Okafor & 2004 & 2010 & \\\hline
\end{tabular}}
\end{examplebox}
\vspace{-3pt}
\begin{examplebox}{(b) Cells paired with headers in DOM order}
\ctx{State › History}: Name: B. Okafor | \hlbad{Term of office: 2004} | \hlbad{Party: 2010}
\end{examplebox}
\vspace{-3pt}
\begin{examplebox}{(c) Row unit from the faithful grid}
\ctx{State › History › Governors} — Name: B. Okafor | \hl{Term of office › Took office: 2004} | \hl{Term of office › Left office: 2010} | \hl{Party: Liberal}
\end{examplebox}
\caption{Faithful grid: span expansion, multi-row headers, and title rows give every cell its correct label.}
\label{fig:grid}
\end{figure}

%% file: figures/block.tex
\begin{figure*}[t]
\centering
\begin{minipage}[t]{0.513\textwidth}
\begin{examplebox}[equal height group=blk]{Coupled context: top-3 chunks of the baseline (excerpts)}
\rdr{\textbf{Q:} What was the strike rate of Harbhajan Singh in the final match of IPL 2015?\hfill Reference: \textbf{200.00}}\\[3pt]
\dimtext{[Chunk 1, same page]} ... | Ravindra Jadeja | not out | 11 | 8 | 1 | 0 | 137.50 | |\\
| Mohit Sharma | not out | 21 | 7 | 1 | 2 | 300.00 | |\\
| Ashish Nehra | |||||||\\
| Extras | (lb 2, w 3, nb 2) | 7 | ||||| ...\\
\dimtext{[Chennai innings rows without their header row; the Mumbai batting rows sit in a stored chunk outside the top three]}\\[3pt]
\dimtext{[Chunk 3, page ``Harbhajan Singh'']} ... The match referee Farokh Engineer found Harbhajan guilty of a level 4.2 offence, banning him from the remainder of the IPL ... Harbhajan had taken five wickets at 16.40 at an economy rate of 8.20 and scored 30 runs at 15.00 in the three matches before his ban ...\\[3pt]
\rdr{\textbf{Reader:} ``Harbhajan Singh did not bat in the final match of IPL 2015 ... the strike rate is not provided.'' \xmark}
\end{examplebox}
\end{minipage}\hfill
\begin{minipage}[t]{0.475\textwidth}
\begin{examplebox}[equal height group=blk]{Decoupled context: \method{} unit block in the third slot}
\ctx{2015 Indian Premier League final › Scorecard › Mumbai Indians innings}\\
\ctx{Player | Status | Runs | Balls | 4s | 6s | Strike rate}\\
Lendl Simmons | b Smith | 68 | 45 | 8 | 3 | 151.11\\
Rohit Sharma* | c Jadeja b Bravo | 50 | 26 | 6 | 2 | 192.30\\
\hl{Harbhajan Singh | not out | 6 | 3 | 0 | 1 | 200.00}\\[3pt]
\ctx{2015 Indian Premier League final › Scorecard › Mumbai Indians bowling}\\
\ctx{Bowler | Overs | Maidens | Runs | Wickets | Econ | Wides | NBs}\\
Harbhajan Singh | 4 | 0 | 34 | 2 | 8.50 | 1 | 0\\[3pt]
\ctx{2013 Indian Premier League final › Scorecard › Mumbai Indians innings}\\
\ctx{Player | Status | Runs | Balls | 4s | 6s | Strike rate}\\
Harbhajan Singh | c Hussey b Bravo | 14 | 8 | 3 | 0 | 175.00\\[3pt]
\rdr{\textbf{Reader:} ``Harbhajan Singh's strike rate in the 2015 IPL final was 200.00 ... listed with 6 runs off 3 balls.'' \cmark}
\end{examplebox}
\end{minipage}
\caption{A packed unit block from \sqa{} (abridged). Rows are scored with their full context, but each group prints its context path and column headers once (green), then its selected rows.}
\label{fig:block}
\end{figure*}

%% file: sections/experiments.tex
\section{Experiments}
\label{sec:experiments}

\ifarxiv
We organize the evaluation around five questions. Does \method{} help pipelines that differ in index, modality, retriever, and reader (\S\ref{sec:main})? Does it beat existing ways of improving retrieved context under a matched budget (\S\ref{sec:baselines})? Does the gain survive a larger reader budget (\S\ref{sec:budget})? Does it carry over to agents that search several times (\S\ref{sec:agentic})? Which components produce it (\S\ref{sec:analysis})?
\fi

\subsection{Setup}
\label{sec:setup}

\para{Datasets.} All pipelines index the August 2025 English Wikipedia snapshot~\cite{kiwix2025zim} (7,060,072 articles) corpus. For evaluation, we use the 3,336 validation questions of Natural Questions (\nq)~\cite{kwiatkowski2019natural} similar to PixelRAG~\cite{wang2026pixelrag}, 1,068 development questions of \nqt~\cite{herzig2021nqtables}, and the 3,447 \sqa questions~\cite{wei2024simpleqa} whose cited source is an English Wikipedia article, totaling 7,851 questions. \nqt targets tabular evidence, and \sqa targets long-tail facts. The agentic evaluation adds two multi-hop benchmarks: all 824 questions of FRAMES~\cite{krishna2025frames}, which need 2 to 15 pages each, and 1,000 answerable MuSiQue questions~\cite{trivedi2022musique}.

\para{Pipelines.} We evaluate three pipelines built by two groups. \emph{Hybrid text + BGE} is our index: we employ Trafilatura~\cite{barbaresi2021trafilatura} for content extraction. Passages are cut into 15.5 million chunks of up to 1,024 tokens, with Qwen3-Embedding-0.6B dense retrieval~\cite{zhang2025qwen3,johnson2021faiss}, BM25~\cite{robertson2009bm25,yang2017anserini} fused by reciprocal rank fusion~\cite{cormack2009rrf}, and BGE-reranker-v2-m3~\cite{chen2024bgem3} reranking over the top 50. \emph{PixelRAG text + BGE} retrieves from the text index released by PixelRAG authors, 15.7 million chunks embedded with Qwen3-VL-Embedding-2B, and reranks using the same reranker. \emph{PixelRAG screenshot + Qwen3-VL} retrieves from the released fine-tuned screenshot index (26.3 million tiles) and reranks with Qwen3-VL-Reranker-2B~\cite{li2026qwenvlretrieval}. Each pipeline passes its top three items to the reader.

\para{Reader and grading.} The default reader is Qwen3.5-4B~\cite{qwen2026qwen35} with greedy decoding and thinking disabled; we also run Gemma 4 E4B and Gemma 4 31B~\cite{google2026gemma4}. GPT-5.6 Luna grades each answer against the reference with the \sqa grading template~\cite{wei2024simpleqa}, and accuracy is the share of answers graded as correct.

\para{\method{} configuration.} One configuration serves every experiment: anchor pages from the top $m=10$ items, a lexical prefilter of $N=100$ units, BGE-reranker-v2-m3 as the scorer, table-grouped packing, and one unit block in the budget of the third evidence item. We fixed the anchor and prefilter sizes before the first evaluation and never tuned the configuration per pipeline, reader, or agent; Appendix~\ref{app:ablation} shows that accuracy is insensitive to these choices. 

\ifarxiv
\para{Statistics.} Every comparison is paired on the same questions. We report exact McNemar tests~\cite{mcnemar1947} and call a difference significant at $p<0.05$; confidence intervals in the appendix come from a bootstrap that resamples source pages. Pooled results sum the paired counts of the three datasets. The study comprises more than one million graded answers, and each comparison reruns its reference condition in the same software environment.
\fi

\subsection{One Layer, Many Pipelines}
\label{sec:main}

\input{tables/main}

Table~\ref{tab:main} applies the same layer to twelve pipeline configurations.

\para{Index and modality.} \method{} raises pooled accuracy by 5.0 points on our hybrid text pipeline, 6.2 points on the released PixelRAG text pipeline, and 5.6 points on the PixelRAG screenshot pipeline, under the same context budget.
The gain is significant on every dataset and pipeline and stays between 4.6 and 5.8 points under four robustness checks (Appendix~\ref{app:robust}): a reader output limit of 16K instead of 200 tokens, a grader that also accepts answers updated since the dataset was built, only questions whose source page supports the reference answer, and only questions that our earlier pipeline experiments did not use. It is largest on \sqa (6.9 to 8.2 points), where long-tail facts often sit in infoboxes and tables, and it reaches 4.4 to 6.3 points on \nqt.
The pipelines start 3.0 points apart and end 1.9 points apart: \method{} adds the most where the pipeline's own chunks miss the most.
Tiles already show tables as the page renders them, but replacing the third tile with a unit block still adds 5.6 points, so the bottleneck that \method{} removes is the unit of delivery in both modalities. We show in Appendix~\ref{app:pixel} that even when rendered as an image tile, our unit block still has noticeable gains.

\para{Retriever.} Six alternative first stages change the baseline by $-1.0$ to $+0.9$ points, and \method{} adds 4.9 to 5.5 points to each of them. The gain does not shrink on the stronger retrievers: it is 5.1 points on top of Query2doc, the best of them, against 5.0 on the default. Retrieval quality and delivery quality improve the pipeline independently, and the combination Query2doc + \method{} reaches 76.7\%, the highest single-shot accuracy in our study.

\para{Reader.} The gain transfers to a second reader family. Gemma 4 E4B, matched in size to the default reader, gains 4.9 points. Gemma 4 31B is eight times larger and improves the baseline by 1.3 points; \method{} adds 4.7 points on top of it for text and 5.9 for screenshots.

\subsection{Comparison with Alternatives}
\label{sec:baselines}

\input{tables/baselines}

Table~\ref{tab:baselines} compares \method{} with fourteen methods on the hybrid text pipeline, with the same questions, reader, and grader, and with every reader context matched to the token count of the top-3 chunks except where the table notes otherwise.

\para{Retrieval-side methods.} RM3 pseudo-relevance feedback~\cite{abduljaleel2004rm3}, the learned sparse retriever SPLADE-v3~\cite{lassance2024spladev3}, the LLM-based query expansions HyDE~\cite{gao2023hyde} and Query2doc~\cite{wang2023query2doc}, late chunking with jina-embeddings-v3~\cite{gunther2024late,sturua2024jina}, and two expansions that append generated queries to every chunk before BM25 indexing, Doc2Query~\cite{nogueira2019doctttttquery} and Doc2Query-{}-~\cite{gospodinov2023doc2query}, each replace or expand one side of first-stage retrieval. None moves pooled accuracy by more than about one point, and \method{} beats all seven by 4.1 to 6.1 points: they reorder chunks, and the reader still reads 1,024-token windows.

\para{Context-enrichment methods.} Small-to-big retrieval~\cite{liu2022llamaindex} ranks 512-token chunks and reads the 1,024-token windows around them, and it ties the baseline. RAPTOR~\cite{sarthi2024raptor} builds a tree of LLM summaries over each anchor page and fills the third slot with tree nodes selected by our cross-encoder; it gains 1.2 points and trails \method{} by 3.8. Contextual retrieval~\cite{anthropic2024contextual} prepends an LLM-written context that situates each chunk in its page; applied to the retrieved candidates before reranking and reading, it gains 0.8 points, all on \sqa, and trails \method{} by 4.3. HtmlRAG~\cite{tan2025htmlrag} fills the third slot with the HTML of the anchor pages, cleaned and pruned by its released two-stage pruner. It ties the top-3 chunks, and with our cross-encoder as its first-stage block ranker (HtmlRAG-CE) it scores 0.7 points below them. \method{} leads both variants by at least 5.4 points, significantly on every dataset, and by 5.8 points at HtmlRAG's own design point of a 4K-token context (Appendix~\ref{app:htmlrag}). HtmlRAG was designed for a handful of search results, and pruning ten full Wikipedia pages into one context leaves fragments without the headers and section paths that \method{} attaches.

\para{Structure-aware chunkers.} We run the rule-based document parsers Unstructured~\cite{unstructured} and Docling~\cite{auer2024docling} as drop-in unit builders inside our delivery layer: the same anchor pages, prefilter, cross-encoder, and slot, with their chunks in place of our units. Both improve on the top-3 chunks, so the delivery layer is useful beyond one parser, and their ranking follows principle P2: Unstructured flattens tables into cell text without headers and gains 2.2 points, while Docling serializes each table row with its headers under the heading path and gains 4.3 points. \method{} is the most accurate builder, 0.8 points above Docling pooled ($p=0.013$), and it parses a page 26 times faster than Unstructured and over 500 times faster than Docling (Appendix~\ref{app:cost}). The gap between the builders widens inside a search agent (\S\ref{sec:agentic}).

\subsection{Accuracy Under a Growing Reader Budget}
\label{sec:budget}

\input{figures/budget}

A natural objection to any context-selection method is that modern readers accept long inputs, so one could pass more chunks instead. Figure~\ref{fig:budget} varies evidence budgets from 2K to 12K tokens. For text, we compare chunks only against three allocations: a 1K-token unit block with chunks in the rest, half units and half chunks, and units only. For screenshots, we compare the top $k$ tiles against the top $k-1$ tiles plus a unit block in the cost of one tile, for $k$ from 2 to 14.

\para{Units beat more context.} \method{} beats the chunks-only context in all 51 pooled comparisons across three chunkings, six budgets, and three allocations by 1.8 to 6.7 points. The two mixed allocations are significant in every dataset in all 42 of their comparisons.
On screenshots, the unit block wins at every budget on every dataset. Extra context helps the baselines slowly: quadrupling the budget from 3K to 12K adds 2.1 points for 1,024-token chunks, 0.9 points for 512-token chunks, and 2.2 points for tiles.

\para{The smallest \method{} context beats the largest baseline.} With 1,024-token chunks, 3K tokens with \method{} reach 76.1\%, and 12K tokens of chunks reach 73.2\%. On the screenshot pipeline, one tile plus a unit block at 2K tokens beats fourteen tiles at 12K by 3.4 points ($p<10^{-16}$) with one sixth of the reader tokens. \method{} therefore reduces reader cost and raises accuracy at once.

\para{Chunk size stops mattering.} Without units, 1,024-token chunks beat 512-token chunks by 1.4 to 2.7 points from 3K upward. With a unit block, the two chunk sizes stay within 0.6 points of each other through 8K. Units provide the fine-grained evidence, and the chunks only need to provide prose, so the pipeline no longer depends on a carefully tuned chunk size.

\para{Units complement, not replace.} The mixed allocations beat units only at every budget by 0.6 to 1.9 points, and at the three-tile budget, one tile plus units is the best screenshot configuration (75.9\%). The pipeline's top items carry running prose that rows and list items cannot replace, which is why \method{} keeps them.

\input{tables/agent}

\subsection{Agentic Search}
\label{sec:agentic}

Agentic RAG lets the model write its own queries, read the results, and search again~\cite{yao2023react,asai2024selfrag,jin2025searchr1}. Each search is a small RAG call, so the delivery layer applies to each call. We evaluate an agent that searches several times per question on multi-hop benchmarks; the arms differ only in the last document of every tool result.

\para{Multi-hop reasoning agent.} Qwen3.5-4B with thinking enabled calls a \texttt{search} tool up to eight times over the hybrid text pipeline and averages 2.1 to 2.8 searches per question across arms on FRAMES and MuSiQue. Table~\ref{tab:agent} shows that \method{} raises FRAMES accuracy from 59.2\% to 64.4\%. The gain concentrates where structured evidence decides the answer: $+9.3$ points on tabular-reasoning questions, $+7.8$ on numerical reasoning, and $+13.1$ on post-processing questions (details in Appendix~\ref{app:agent}). MuSiQue composes its questions from single paragraphs, mostly lead-section prose that the top-2 chunks already carry, and it gains 1.1 points.

We provide two more controls to isolate the cause. \textit{A fourth chunk per search}, about a third more tokens, does not bring benefits ($-0.05$ points). \textit{Docling chunks} in the same slot lower accuracy by 0.9 points and trail \method{} by 3.8 points pooled and by 6.8 points on FRAMES.
The agent also behaves differently: with \method{} it searches less often than with chunks (2.19 against 2.41 searches) and finishes with the shortest context (11.7K tokens vs. 12.8K), whereas with Docling it issues a third search on 29\% of FRAMES questions against 16\%. Compact, self-describing evidence lets the agent stop earlier while achieving better performance.

\subsection{Analysis}
\label{sec:analysis}

\input{tables/delivery}

\para{Ablations.}
\label{sec:ablation}
We conduct ablations to show that completion brings the gain: bare units, the same rows and paragraphs without titles, sections, and headers add only 1.8 points, while completed units add 4.6.
We also test different row constructions and serializations, which stay within 0.34 points of each other: for example, joint ranking of chunks and units loses 0.6 points, units only lose 1.6, and a second unit slot, other anchor counts, or prefilter sizes change accuracy by at most 0.3 points (details in Appendix~\ref{app:ablation}).

\para{\method{} gains where the evidence went missing.} Table~\ref{tab:delivery} groups the 5,883 source-supported questions of \S\ref{sec:trace} by where the top-3 pipeline loses the quoted evidence, matched tolerantly against the exact reader inputs: the source page is not among the ten anchor pages, extraction or chunking breaks the evidence, or a stored chunk holds it but misses the reader. \method{} raises evidence delivery from 72.1\% to 81.9\% of these questions. The two groups that the delivery layer targets hold 21\% of the questions and 225 of the 292 net gains (77\%). Questions whose evidence sits only in tables or infoboxes gain 13.8 points, and prose-only questions gain 2.4. On the oracle questions (Table~\ref{tab:oracle}), \method{} raises accuracy from 49.1\% to 65.0\%, closing 54\% of the oracle's gain overall and 74\% where the evidence was stored but not delivered.

\para{Remaining failures.} Of \method{}'s 893 errors on source-supported questions, 425 occur with the quoted evidence already in the reader input, which would improve with a stronger reader; an audit of 150 failures attributes most \nq and \nqt failures to outdated references and grading errors (Appendix~\ref{app:robust}). On the oracle questions, the oracle stays 13.4 points ahead (Appendix~\ref{app:trace}). Of the 96 answers in this gap, 61 lie on pages outside the anchor set, where a stronger retriever adds to \method{} (\S\ref{sec:main}), and 35 lie within anchor pages, split between the lexical prefilter (20) and selection and packing (15). A larger prefilter changes accuracy by at most 0.3 points, so recovering these units calls for semantic matching in the prefilter.

%% file: tables/main.tex
\begin{table*}[t]
\caption{Accuracy (\%) with the pipeline's top-3 evidence $\rightarrow$ with \method{} (top-2 + unit block, same budget), and the gain.}
\label{tab:main}
\centering
\setlength{\tabcolsep}{4pt}
\newcommand{\cellx}[3]{#1 $\rightarrow$ \textbf{#2} \gain{#3}}
\begin{tabular}{@{}llllll@{}}
\toprule
\headrow \textbf{Pipeline} & \textbf{Reader} & \textbf{\nq} (3,336) & \textbf{\nqt} (1,068) & \textbf{\sqa} (3,447) & \textbf{Pooled} \\
\midrule
\grouprow{6}{Retrieval pipeline and modality}
Hybrid text + BGE & Qwen3.5-4B & \cellx{64.6}{67.8}{+3.2} & \cellx{47.9}{52.3}{+4.4} & \cellx{83.6}{90.6}{+7.0} & \cellx{70.7}{75.7}{+5.0} \\
PixelRAG text + BGE & Qwen3.5-4B & \cellx{62.2}{66.3}{+4.1} & \cellx{45.9}{52.2}{+6.3} & \cellx{79.8}{88.0}{+8.2} & \cellx{67.7}{73.9}{+6.2} \\
PixelRAG screenshot + VL & Qwen3.5-4B & \cellx{63.1}{67.2}{+4.1} & \cellx{45.8}{51.8}{+6.0} & \cellx{83.3}{90.1}{+6.9} & \cellx{69.6}{75.2}{+5.6} \\
\grouprow{6}{First-stage retriever of the hybrid text pipeline}
BM25 + RM3 feedback & Qwen3.5-4B & \cellx{64.4}{67.6}{+3.2} & \cellx{47.3}{51.6}{+4.3} & \cellx{82.4}{89.8}{+7.5} & \cellx{70.0}{75.2}{+5.2} \\
SPLADE-v3 for BM25 & Qwen3.5-4B & \cellx{65.0}{68.6}{+3.5} & \cellx{49.0}{52.6}{+3.7} & \cellx{80.6}{88.5}{+7.9} & \cellx{69.7}{75.2}{+5.5} \\
HyDE & Qwen3.5-4B & \cellx{65.6}{68.7}{+3.1} & \cellx{48.5}{53.7}{+5.2} & \cellx{83.7}{90.9}{+7.2} & \cellx{71.2}{76.4}{+5.2} \\
Query2doc & Qwen3.5-4B & \cellx{66.4}{69.4}{+3.0} & \cellx{48.3}{53.1}{+4.8} & \cellx{83.9}{91.0}{+7.1} & \cellx{71.6}{76.7}{+5.1} \\
Doc2Query for BM25 & Qwen3.5-4B & \cellx{64.8}{68.1}{+3.3} & \cellx{48.6}{52.3}{+3.7} & \cellx{83.7}{90.5}{+6.8} & \cellx{70.9}{75.8}{+4.9} \\
Doc2Query-{}- for BM25 & Qwen3.5-4B & \cellx{64.8}{68.7}{+3.8} & \cellx{47.6}{52.3}{+4.8} & \cellx{83.5}{90.9}{+7.4} & \cellx{70.7}{76.2}{+5.5} \\
\grouprow{6}{Reader model}
Hybrid text + BGE & Gemma 4 E4B & \cellx{64.5}{67.2}{+2.7} & \cellx{46.5}{52.3}{+5.8} & \cellx{84.3}{91.0}{+6.7} & \cellx{70.7}{75.6}{+4.9} \\
Hybrid text + BGE & Gemma 4 31B & \cellx{65.2}{68.4}{+3.2} & \cellx{48.7}{54.3}{+5.6} & \cellx{85.8}{91.6}{+5.9} & \cellx{72.0}{76.7}{+4.7} \\
PixelRAG screenshot + VL & Gemma 4 31B & \cellx{65.8}{70.2}{+4.4} & \cellx{49.3}{55.4}{+6.1} & \cellx{83.8}{91.0}{+7.2} & \cellx{71.5}{77.4}{+5.9} \\
\bottomrule
\end{tabular}
\end{table*}

%% file: tables/baselines.tex
\begin{table}[t]
\caption{Matched-budget comparison on the hybrid text pipeline (7,851 questions, 3K evidence tokens). $\Delta$: pooled change vs.\ the top-3 chunks; gray: not significant. $^\ddagger$Applied to the retrieved candidates; about 170 more reader tokens.}
\label{tab:baselines}
\centering
\setlength{\tabcolsep}{2.6pt}
\begin{tabular}{@{}lccccr@{}}
\toprule
\headrow \textbf{Method} & \textbf{\nq} & \textbf{NQ-Tab.} & \textbf{\sqa} & \textbf{Pooled} & \textbf{$\Delta$} \\
\midrule
Top-3 chunks & 64.63 & 47.94 & 83.64 & 70.70 & \\
\grouprow{6}{Retrieval-side methods}
BM25 + RM3~\cite{abduljaleel2004rm3} & 64.39 & 47.28 & 82.36 & 69.95 & \loss{$-$0.75} \\
SPLADE-v3~\cite{lassance2024spladev3} & 65.02 & 48.97 & 80.62 & 69.69 & \loss{$-$1.02} \\
HyDE~\cite{gao2023hyde} & 65.62 & 48.50 & 83.72 & 71.24 & \gain{+0.53} \\
Query2doc~\cite{wang2023query2doc} & 66.37 & 48.31 & 83.87 & 71.60 & \gain{+0.89} \\
Late chunking~\cite{gunther2024late} & 64.96 & 47.85 & 83.46 & 70.76 & \nsig{+0.05} \\
Doc2Query~\cite{nogueira2019doctttttquery} & 64.84 & 48.60 & 83.70 & 70.91 & \nsig{+0.20} \\
Doc2Query-{}-~\cite{gospodinov2023doc2query} & 64.84 & 47.57 & 83.49 & 70.68 & \nsig{$-$0.03} \\
\grouprow{6}{Context-enrichment methods}
Small-to-big~\cite{liu2022llamaindex} & 64.33 & 47.38 & 83.72 & 70.54 & \nsig{$-$0.17} \\
RAPTOR~\cite{sarthi2024raptor} & 65.92 & 48.41 & 85.00 & 71.91 & \gain{+1.21} \\
Contextual retr.$^\ddagger$~\cite{anthropic2024contextual} & 64.63 & 47.00 & 85.70 & 71.48 & \gain{+0.78} \\
HtmlRAG~\cite{tan2025htmlrag} & 65.08 & 47.57 & 82.39 & 70.30 & \nsig{$-$0.41} \\
HtmlRAG-CE & 64.69 & 46.91 & 82.33 & 70.02 & \loss{$-$0.69} \\
\grouprow{6}{Structure-aware chunkers as our unit builder}
Unstructured~\cite{unstructured} & 66.01 & 50.66 & 86.36 & 72.86 & \gain{+2.16} \\
Docling~\cite{auer2024docling} & \best{67.84} & 51.03 & 89.30 & 74.97 & \gain{+4.27} \\
\midrule
\oursrow \method{} & \best{67.84} & \best{52.34} & \best{90.63} & \best{75.74} & \gain{\textbf{+5.03}} \\
\oursrow Query2doc + \method{} & \textit{69.39} & \textit{53.09} & \textit{90.98} & \textit{76.65} & \textbf{+5.95} \\
\bottomrule
\end{tabular}
\end{table}

%% file: figures/budget.tex
\begin{figure*}[t]
\centering
\begin{tikzpicture}
\begin{groupplot}[
  group style={group size=3 by 1, horizontal sep=1.25cm},
  paperplot, width=6.2cm, height=4.1cm,
  symbolic x coords={2K,3K,4K,6K,8K,12K}, xtick=data,
  ymin=67.5, ymax=78, ytick={68,70,72,74,76,78},
  xlabel={Reader evidence budget (tokens)},
  enlarge x limits=0.07,
]
\nextgroupplot[title={(a) Text, 1,024-token chunks}, ylabel={Pooled accuracy (\%)},
  legend to name=budgetlegend, legend columns=4, legend style={column sep=8pt}]
\addplot[base] coordinates {(2K,69.43) (3K,71.09) (4K,71.39) (6K,72.27) (8K,73.06) (12K,73.20)};
\addplot[ours] coordinates {(2K,75.42) (3K,75.84) (4K,76.14) (6K,76.60) (8K,76.26) (12K,76.68)};
\addplot[oursalt] coordinates {(3K,76.12) (4K,76.18) (6K,76.21) (8K,76.22) (12K,77.01)};
\addplot[only] coordinates {(2K,74.39) (3K,74.51) (4K,75.02) (6K,74.75) (8K,74.89) (12K,75.30)};
\draw[cMute, densely dashed, line width=0.6pt] (axis cs:3K,76.12) -| (axis cs:12K,73.20);
\node[font=\figfont\footnotesize, text=cInk, anchor=north east, align=right] at (axis cs:12K,71.4) {3K with \method{}\\[-2pt]$>$ 12K of chunks};
\legend{Chunks or tiles only, {\method{}: 1 unit block (1K tokens)}, {\method{}: half the budget}, Units only}
\nextgroupplot[title={(b) Text, 512-token chunks}]
\addplot[base] coordinates {(2K,69.01) (3K,69.60) (4K,70.03) (6K,70.70) (8K,70.50) (12K,70.48)};
\addplot[ours] coordinates {(2K,75.75) (3K,75.84) (4K,75.99) (6K,76.04) (8K,75.89) (12K,75.60)};
\addplot[oursalt] coordinates {(3K,75.93) (4K,76.13) (6K,76.14) (8K,75.91) (12K,76.17)};
\addplot[only] coordinates {(2K,74.44) (3K,74.54) (4K,74.81) (6K,74.70) (8K,74.36) (12K,75.02)};
\nextgroupplot[title={(c) Screenshot tiles (2 to 14 tiles)}]
\addplot[base] coordinates {(2K,68.22) (3K,69.58) (4K,70.16) (6K,71.04) (8K,71.47) (12K,71.76)};
\addplot[ours] coordinates {(2K,75.18) (3K,75.18) (4K,75.20) (6K,75.44) (8K,75.57) (12K,75.32)};
\draw[cMute, densely dashed, line width=0.6pt] (axis cs:2K,75.18) -| (axis cs:12K,71.76);
\node[font=\figfont\footnotesize, text=cInk, anchor=north east, align=right] at (axis cs:12K,74.8) {1 tile + units\\[-2pt]$>$ 14 tiles};
\end{groupplot}
\end{tikzpicture}\\[-2pt]
\pgfplotslegendfromname{budgetlegend}
\caption{Accuracy against reader budget on 7,851 questions. In (c), one tile costs 864 reader tokens, as does the unit block.}
\label{fig:budget}
\end{figure*}
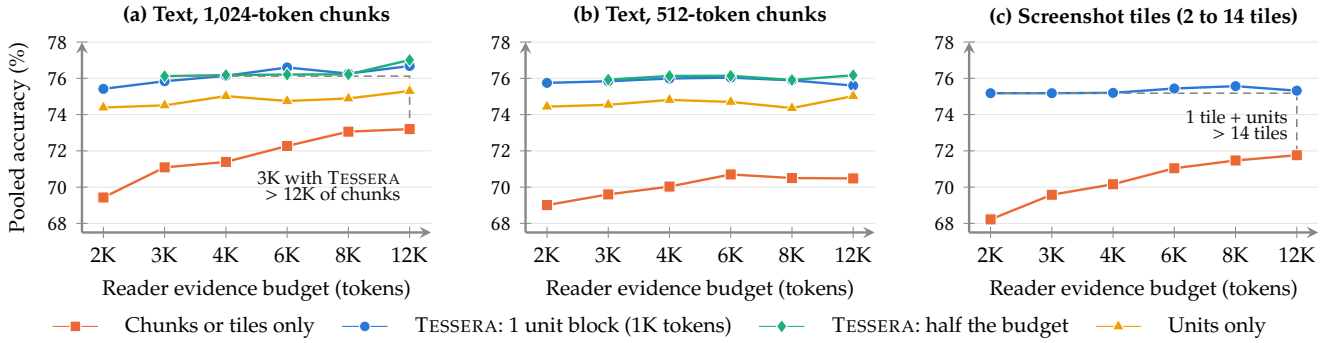

%% file: tables/agent.tex
\begin{table}[t]
\caption{\method{} inside a multi-step search agent (Qwen3.5-4B with thinking, up to 8 searches). \emph{Srch.}: mean searches per question; \emph{Ctx.}: mean final prompt length (K tokens).}
\label{tab:agent}
\centering
\setlength{\tabcolsep}{2.8pt}
\begin{tabular}{@{}lccccc@{}}
\toprule
\headrow \textbf{Search result} & \textbf{FRAMES} & \textbf{MuSiQue} & \textbf{Pooled} & \textbf{Srch.} & \textbf{Ctx.} \\
\midrule
Top-3 chunks & 59.22 & 32.40 & 44.52 & 2.41 & 12.8 \\
Top-4 chunks & 59.22 & 32.30 & 44.46 & 2.18 & 14.5 \\
Top-2 + Docling & 57.65 & 32.10 & 43.64 & 2.65 & 13.6 \\
\oursrow Top-2 + \method{} & \best{64.44} & \best{33.50} & \best{47.48} & 2.19 & \best{11.7} \\
\oursrow \quad gain vs.\ top-3 & \gain{+5.22} & \nsig{+1.10} & \gain{+2.96} & & \\
\bottomrule
\end{tabular}

\end{table}

%% file: tables/delivery.tex
\begin{table}[t]
\caption{Accuracy (\%) on the source-supported questions, grouped by where the top-3 pipeline loses the quoted evidence. Net: questions gained minus questions lost.}
\label{tab:delivery}
\centering
\setlength{\tabcolsep}{4pt}
\begin{tabular}{@{}lrccr@{}}
\toprule
\headrow \textbf{Evidence in the top-3 pipeline} & \textbf{$n$} & \textbf{Top-3} & \textbf{\method{}} & \textbf{Net} \\
\midrule
Page not anchored & 421 & 39.9 & 43.7 & +16 \\
Lost in extraction or chunking & 390 & 60.8 & \best{82.3} & \gain{+84} \\
Stored, not delivered & 832 & 54.0 & \best{70.9} & \gain{+141} \\
Delivered & 4,240 & 90.7 & 91.9 & +51 \\
\midrule
All & 5,883 & 79.9 & \best{84.8} & \gain{+292} \\
\bottomrule
\end{tabular}
\end{table}

%% file: sections/related.tex
\section{Related Work}
\label{sec:related}

\ifarxiv
We review prior work along the pipeline, from indexing to reading.
\fi

\para{Web content and tables in RAG.} Web RAG systems commonly reduce a page to main-content text before indexing~\cite{barbaresi2021trafilatura}. Learned extractors~\cite{wang2025readerlm,liu2025dripper} preserve more structure at a much higher corpus-processing cost, and HtmlRAG~\cite{tan2025htmlrag} keeps pruned HTML in the reader's context. Table QA serializes a given table for a language model~\cite{sui2024table} or retrieves tables from a table corpus~\cite{herzig2021nqtables}, treating each table as standalone. \method{} shares the premise that page structure carries information but converts it into explicit context on small units, with no markup tokens or trained models, and selects table rows, merged cells resolved from the HTML, under a budget shared with the surrounding prose.

\para{Visual document retrieval.} DSE~\cite{ma2024dse}, ColPali~\cite{faysse2025colpali}, VisRAG~\cite{yu2025visrag}, and M3DocRAG~\cite{cho2024m3docrag} retrieve rendered pages with vision-language encoders, and PixelRAG~\cite{wang2026pixelrag} scales the approach to a full Wikipedia snapshot, motivated by the rich layout information present in webpages. A screenshot tile is a fixed window like a chunk, so \method{} complements this line: it uses the top-ranked tiles only to find which pages to open, and builds its units from their HTML (\S\ref{sec:main}).

\para{Retrieval granularity and chunk context.} Proposition-level indexing~\cite{chen2024densex}, small-to-big retrieval~\cite{liu2022llamaindex}, RAPTOR~\cite{sarthi2024raptor}, late chunking~\cite{gunther2024late}, contextual retrieval~\cite{anthropic2024contextual}, and structure-aware parsers~\cite{auer2024docling,unstructured} all adjust what a chunk is when the corpus is indexed. They pay their cost for every page, and most operate on extracted text, which has already lost the structured evidence. \method{} acts at query time on the HTML of the pages a query retrieves and is independent of the index granularity. Query and document expansion~\cite{abduljaleel2004rm3,lassance2024spladev3,gao2023hyde,wang2023query2doc,nogueira2019doctttttquery,gospodinov2023doc2query} and cross-encoder reranking~\cite{chen2024bgem3} address earlier stages, and \S\ref{sec:main} shows that their gains add to ours.

\para{Context compression.} Post-retrieval refiners such as RECOMP~\cite{xu2024recomp} and LongLLMLingua~\cite{jiang2024longllmlingua} shorten retrieved passages to reduce costs and distractions. \method{} instead selects complete structured units from the pages behind them and is orthogonal to such refiners.

\para{Agentic RAG.} ReAct-style agents~\cite{yao2023react}, Self-RAG~\cite{asai2024selfrag}, and RL-trained agents such as Search-R1~\cite{jin2025searchr1} optimize when and what to retrieve but take the tool's output format as given; we show that this format shapes the policy: with self-describing evidence, an agent searches less and answers more questions correctly (\S\ref{sec:agentic}).

%% file: sections/conclusion.tex
\section{Conclusion}
\label{sec:conclusion}

Web RAG pipelines index, rank, and read the same object, and our source-to-reader trace shows the cost of that coupling: answer evidence that the corpus contains fails to reach the reader, in text pipelines and screenshot pipelines alike. \method{} decouples delivery from retrieval. It reopens the pages a pipeline retrieves, rebuilds them into structure-completed \units{}, and spends one evidence slot on the units the question needs. The layer is training-free and index-free. \ifarxiv It adds 5 to 6 accuracy points to three pipelines built on two modalities by two groups, holds its gain across readers, retrievers, chunk sizes, and budgets, outperforms\fi{} fourteen alternative methods at matched cost, and makes a multi-step search agent up to 5.2 points more accurate while issuing fewer searches and reading a shorter context.

\para{Scope and future work.} Our experiments use English Wikipedia, whose pages are rich in tables and infoboxes; the open Web adds layout tables, client-side rendering, and noisier markup. Extending the unit builder to them requires a specific larger corpus and is our next step for future work.
The layer requires only access to page HTML at query time, without additional training or models.
Two further directions follow from our analysis: wider anchoring and a semantic prefilter, since most of the remaining oracle gap lies on unanchored pages and in units that the lexical prefilter misses; and search agents trained to issue queries that target unit blocks.


%% file: sections/appendix.tex
\section{Implementation and Reproducibility}
\label{app:impl}

All runs use a consistent Python environment (vLLM 0.25.1) on NVIDIA A100 GPUs. We will release the code, run manifests, unit builder, and evaluation records.

\para{Pipeline details.} Beyond the setup of \S\ref{sec:setup}, the hybrid text pipeline extracts text with Trafilatura 2.0.0, retrieves the top 100 candidates each from a FAISS IVFFlat index (4,096 lists, 128 probes) and from Anserini BM25 ($k_1{=}1.2$, $b{=}0.6$), and fuses them by reciprocal rank fusion with constant 60. The 512-token chunking yields 26.8 million chunks. The reader's output limit is 200 tokens, and the \method{} scorer truncates each question--unit pair at 512 tokens.

\para{Baselines.} HtmlRAG uses the released \texttt{htmlrag} package with its HTML cleaning, block tree, BGE-large first-stage ranker over 256-word blocks, and HTML-Pruner-Llama-1B over 128-word blocks, applied to the same anchor pages. RAPTOR uses the reference settings (100-token leaves, soft clustering, 100-token summaries) with Qwen3.5-4B as the summarizer, one tree per anchor page. Late chunking embeds all 15.5 million chunks with jina-embeddings-v3, once alone and once inside 8,192-token article windows, and each index replaces the dense retriever. Query2doc and HyDE use Qwen3.5-4B as a zero-shot generator with the instructions of the original papers. RM3 uses the Anserini defaults. SPLADE-v3 replaces BM25. Doc2Query samples 10 queries with doc2query-t5-base-msmarco for each 511-token segment of every chunk and appends them before BM25 indexing; Doc2Query-{}- keeps the 30\% of queries that monoELECTRA scores highest against their segment. Contextual retrieval uses the published prompt with Qwen3.5-4B over a window of at most 8,192 tokens of the chunk's article; we contextualize the 323,152 chunks among the top-50 candidates of the evaluation questions and do not re-index the corpus. Docling uses its HTML backend and hybrid chunker (256 tokens, contextualized headings); Unstructured uses \texttt{partition\_html} with \texttt{chunk\_by\_title}.

\section{Evidence Trace Details}
\label{app:trace}

Table~\ref{tab:gapfull} accounts for the oracle gap of \S\ref{sec:analysis}. Panel (a) groups the 715 oracle questions by where the top-3 pipeline loses the evidence: \method{} closes 98 of the 133 answers (74\%) where the evidence was stored but not delivered. Oracle eligibility uses exact matching, and the stage analysis uses tolerant matching, under which 93 oracle questions already have their evidence in the reader input. Panel (b) groups all 5,883 source-supported questions by where \method{} loses the evidence. Overall, the oracle leads \method{} by 13.4 points (95\% CI 10.2 to 16.7, $p=5.4\times10^{-15}$). Table~\ref{tab:control} adds the same-page control to the oracle intervention: an arbitrary chunk of similar length from the page of the evidence chunk leaves accuracy at the baseline level overall, and the evidence chunk beats it by 16.7 to 50.3 points on every dataset.

\begin{table}[h]
\caption{Oracle gap accounting. Gap: oracle minus \method{} correct answers on the oracle questions.}
\label{tab:gapfull}
\centering
\setlength{\tabcolsep}{3.5pt}
\resizebox{\columnwidth}{!}{%
\begin{tabular}{@{}lrrrr@{}}
\toprule
\multicolumn{5}{@{}l}{\textit{(a) Oracle questions, by top-3 evidence stage}} \\
\headrow \textbf{Evidence in the top-3 pipeline} & \textbf{$n$} & \textbf{Oracle gain} & \textbf{Closed} & \textbf{Gap} \\
\midrule
Page not anchored & 173 & 69 & 9 & 60 \\
Stored, not delivered & 449 & 133 & 98 & 35 \\
Delivered & 93 & 8 & 7 & 1 \\
All & 715 & 210 & 114 & 96 \\
\midrule
\multicolumn{5}{@{}l}{\textit{(b) Source-supported questions, by \method{} evidence stage}} \\
\headrow \textbf{Evidence in \method{}} & \textbf{$n$} & \textbf{Errors} & & \textbf{Gap} \\
\midrule
Page outside top-100 & 182 & 107 & & 27 \\
Page outside reranker pool & 79 & 51 & & 18 \\
Page below anchor cutoff & 123 & 61 & & 16 \\
Not in any unit & 143 & 24 & & 1 \\
Missed by lexical prefilter & 321 & 152 & & 20 \\
Missed by selection and packing & 217 & 73 & & 15 \\
Delivered & 4,818 & 425 & & $-$1 \\
All & 5,883 & 893 & & 96 \\
\bottomrule
\end{tabular}}
\end{table}

\begin{table}[h]
\caption{Oracle intervention with a same-page control (questions that have a control chunk within 20\% of the evidence chunk's length). Entries are correct answers.}
\label{tab:control}
\centering
\setlength{\tabcolsep}{4pt}
\begin{tabular}{@{}lrrrrl@{}}
\toprule
\headrow \textbf{Dataset} & \textbf{$n$} & \textbf{Top-3} & \textbf{Control} & \textbf{Evidence} & \textbf{vs.\ control} \\
\midrule
\nq & 454 & 242 & 265 & 341 & \gain{+16.7} (92/16) \\
\nqt & 56 & 28 & 27 & 41 & \gain{+25.0} (15/1) \\
\sqa & 157 & 69 & 61 & 140 & \gain{+50.3} (79/0) \\
All & 667 & 339 & 353 & 522 & \gain{+25.3} (186/17) \\
\bottomrule
\end{tabular}
\end{table}

\section{Standard Remedies in Detail}
\label{app:remedies}

Table~\ref{tab:remedies} gives the accuracy change of each remedy of \S\ref{sec:remedies}, all applied to the hybrid text pipeline.

\para{Better extraction.} Re-extracting the corpus with Trafilatura's recall-oriented mode and rebuilding the index doubles the coverage of the 449 lost-evidence questions, yet accuracy changes by at most 0.8 points. The recovered text still has to win one of three reader slots as part of a 1,024-token window.

\para{More context around the hits.} Adding the stored neighbors of every hit on the ten best pages to the candidate pool and reranking the enlarged pool brings at most 1.5 points.

\para{Structure as a ranking signal.} Prefixing each chunk with its page title moves the evidence chunk into the reranker's top three for a net 66 of 3,718 questions; the heading path adds three more, and table headers push out more evidence chunks than they bring in. Showing the reader the title-prefixed chunks moves accuracy by less than a point in either direction.

\para{A larger reader budget.} Raising the evidence budget from 3K to 12K tokens lifts pooled accuracy from 71.1\% to 73.2\%: 2.1 points for four times the reader cost.

\para{Screenshots.} Tiles show the tables that the extractor drops, yet the screenshot pipeline shares the ceiling of the text pipeline (Table~\ref{tab:gap}), and raising its context from 3 to 14 tiles adds only 2.2 points.

 \para{Learned extraction.} Beyond Trafilatura's recall-oriented mode, we also tested the LLM extractor MinerU-HTML~\cite{liu2025dripper}. It covers 275 of the 449 lost-evidence questions, yet it classifies infoboxes as boilerplate (97 of its 156 missed table cells are infobox cells) and would take about 80 A100 GPU-days to process the corpus.

\input{tables/remedies}

\section{Ablations}
\label{app:ablation}

\para{Completion brings the gain.} Panel (a) of Table~\ref{tab:ablation} varies the content of the third slot. Bare units, which are the same rows and paragraphs without titles, sections, and headers, add 1.8 points pooled and nothing on \nqt. Completed units add 4.6 points, and they significantly outperform bare units on every dataset. Prefixing the page title to every chunk, the common ``contextual header'' practice, adds 0.5 points. Adding the unit block as a fourth item, at about 900 more tokens, improves the budget-neutral replacement by 0.2 points: \textbf{the content of the slot matters, but its extra tokens do not}.

\para{Faithful grids help tables; serialization is robust.} Panel (b) holds the content fixed and varies construction and packing. The faithful grid corrects the labels of a third of all table rows, and its largest effect is on \nqt ($+1.4$ points over DOM-order pairing). Grouped packing fits a quarter more lines into the slot. The four variants stay within 0.34 points pooled, so the method does not hinge on one serialization; we adopt the grid with table grouping because it is faithful to the source table and the most compact.

\para{The fixed split is the right allocation.} Panel (c) varies the budget split and the pool. Ranking chunks and units jointly by cross-encoder score loses 0.6 points because scores favor long chunks over one-row units and squeeze the unit block out on 13\% of \sqa questions. Replacing all chunks with units loses 1.6 points. A second unit slot, half or double the anchor pages, and half or double the prefilter change accuracy by at most 0.3 points. Adding the sibling rows of the best-scoring row and removing units whose text the chunks already contain do not help either. \textbf{The default configuration sits on a wide plateau.}

\input{tables/ablation}

\section{Complete Budget Results}
\label{app:budget}

Table~\ref{tab:budgetfull} lists every cell of the text budget sweep of \S\ref{sec:budget}, including 512-token chunks with a 64-token overlap, and Table~\ref{tab:pixelbudget} gives the screenshot sweep per dataset. Every gain in Table~\ref{tab:pixelbudget} is significant.

\begin{table}[h]
\caption{Pooled accuracy (\%) by chunking, evidence budget, and allocation. ``+1K units'' gives 1,024 tokens to the unit block and the rest to chunks.}
\label{tab:budgetfull}
\centering
\setlength{\tabcolsep}{4pt}
\begin{tabular}{@{}llcccc@{}}
\toprule
\headrow \textbf{Chunks} & \textbf{Budget} & \textbf{Chunks} & \textbf{+1K units} & \textbf{Half} & \textbf{Units only} \\
\midrule
\multirow{6}{*}{1,024} & 2K & 69.43 & \best{75.42} & -- & 74.39 \\
 & 3K & 71.09 & 75.84 & \best{76.12} & 74.51 \\
 & 4K & 71.39 & 76.14 & \best{76.18} & 75.02 \\
 & 6K & 72.27 & \best{76.60} & 76.21 & 74.75 \\
 & 8K & 73.06 & \best{76.26} & 76.22 & 74.89 \\
 & 12K & 73.20 & 76.68 & \best{77.01} & 75.30 \\
\midrule
\multirow{6}{*}{512} & 2K & 69.01 & \best{75.75} & -- & 74.44 \\
 & 3K & 69.60 & 75.84 & \best{75.93} & 74.54 \\
 & 4K & 70.03 & 75.99 & \best{76.13} & 74.81 \\
 & 6K & 70.70 & 76.04 & \best{76.14} & 74.70 \\
 & 8K & 70.50 & 75.89 & \best{75.91} & 74.36 \\
 & 12K & 70.48 & 75.60 & \best{76.17} & 75.02 \\
\midrule
\multirow{6}{*}{\shortstack[l]{512,\\overlap}} & 2K & 69.51 & \best{75.44} & -- & 74.37 \\
 & 3K & 70.19 & 75.38 & \best{75.85} & 74.58 \\
 & 4K & 70.58 & 75.89 & \best{76.23} & 74.79 \\
 & 6K & 70.96 & 75.93 & \best{76.07} & 74.82 \\
 & 8K & 71.28 & 76.00 & \best{76.27} & 74.65 \\
 & 12K & 71.28 & 76.22 & \best{76.33} & 75.18 \\
\bottomrule
\end{tabular}
\end{table}

\begin{table}[h]
\caption{Screenshot pipeline: top-$k$ tiles $\rightarrow$ top-$(k{-}1)$ tiles + unit block at the same reader cost (accuracy, \%).}
\label{tab:pixelbudget}
\centering
\setlength{\tabcolsep}{3pt}
\begin{tabular}{@{}lrcccl@{}}
\toprule
\headrow \textbf{Budget} & \textbf{$k$} & \textbf{\nq} & \textbf{\nqt} & \textbf{\sqa} & \textbf{Pooled} \\
\midrule
2K & 2 & 61.0 $\rightarrow$ 67.4 & 45.5 $\rightarrow$ 50.6 & 82.3 $\rightarrow$ 90.4 & \gain{+6.95} \\
3K & 3 & 63.1 $\rightarrow$ 67.2 & 45.8 $\rightarrow$ 51.8 & 83.3 $\rightarrow$ 90.1 & \gain{+5.59} \\
4K & 4 & 63.7 $\rightarrow$ 67.0 & 47.4 $\rightarrow$ 50.7 & 83.5 $\rightarrow$ 90.7 & \gain{+5.04} \\
6K & 7 & 64.8 $\rightarrow$ 67.5 & 48.1 $\rightarrow$ 51.0 & 84.1 $\rightarrow$ 90.8 & \gain{+4.41} \\
8K & 9 & 65.3 $\rightarrow$ 67.3 & 48.6 $\rightarrow$ 52.1 & 84.5 $\rightarrow$ 90.8 & \gain{+4.10} \\
12K & 14 & 65.4 $\rightarrow$ 67.1 & 49.3 $\rightarrow$ 52.0 & 84.8 $\rightarrow$ 90.5 & \gain{+3.55} \\
\bottomrule
\end{tabular}
\end{table}

\section{HtmlRAG at Its Design Point}
\label{app:htmlrag}

HtmlRAG targets a 4K-token context. Table~\ref{tab:htmlrag4k} evaluates it there on all 7,851 questions with the same anchor pages as \method{}. Pruned HTML alone loses most answer-bearing text when ten Wikipedia pages are pruned into 4K tokens, and the reader declines to answer 1,305 questions. Placed after the top-2 chunks, either HtmlRAG variant scores below chunks alone. \method{} leads the best HtmlRAG configuration by 5.8 points ($p<10^{-49}$), and \method{} units alone, without any chunk, reach 74.6\%.

\begin{table}[h]
\caption{HtmlRAG and \method{} at a 4,096-token evidence budget (accuracy, \%). \emph{Decl.}: questions the reader declined.}
\label{tab:htmlrag4k}
\centering
\setlength{\tabcolsep}{2.6pt}
\resizebox{\columnwidth}{!}{%
\begin{tabular}{@{}lccccr@{}}
\toprule
\headrow \textbf{Context} & \textbf{\nq} & \textbf{NQ-Tab.} & \textbf{\sqa} & \textbf{Pooled} & \textbf{Decl.} \\
\midrule
Chunks & 65.83 & 48.03 & 84.31 & 71.52 & 118 \\
HtmlRAG & 41.28 & 25.84 & 17.52 & 28.75 & 1,305 \\
HtmlRAG-CE & 42.33 & 26.87 & 21.85 & 31.23 & 1,283 \\
Top-2 + HtmlRAG & 64.36 & 46.72 & 82.56 & 69.95 & 172 \\
Top-2 + HtmlRAG-CE & 64.45 & 48.03 & 83.14 & 70.42 & 174 \\
\method{} units only & 68.08 & 50.66 & 88.22 & 74.55 & 96 \\
\oursrow Top-2 + \method{} & \best{68.59} & \best{53.37} & \best{90.60} & \best{76.18} & \best{80} \\
\bottomrule
\end{tabular}}
\end{table}

\section{Agentic Search Details}
\label{app:agent}

\para{Protocol.} The reasoning agent is Qwen3.5-4B served with its own chat template and function-calling format, thinking enabled, the model card's sampling settings with a fixed seed, and at most eight calls to a single \texttt{search(query)} tool. Each call runs the hybrid text pipeline on the agent's query and returns three documents. Table~\ref{tab:frames} breaks down the FRAMES result by annotated reasoning type.

\begin{table}[h]
\caption{FRAMES accuracy (\%) of the reasoning agent by reasoning type; a question can carry several. G/L: questions gained and lost by \method{} against the top-3 chunks.}
\label{tab:frames}
\centering
\setlength{\tabcolsep}{2.6pt}
\resizebox{\columnwidth}{!}{%
\begin{tabular}{@{}lrccccl@{}}
\toprule
\headrow \textbf{Type} & \textbf{$n$} & \textbf{Chunks} & \textbf{Docling} & \textbf{\method{}} & \textbf{G/L} & \textbf{$p$} \\
\midrule
Tabular & 236 & 50.8 & 51.7 & \best{60.2} & 44/22 & 0.009 \\
Numerical & 293 & 53.2 & 52.9 & \best{61.1} & 51/28 & 0.013 \\
Post-proc. & 107 & 47.7 & 49.5 & \best{60.7} & 21/7 & 0.013 \\
Temporal & 278 & 57.6 & 55.0 & \best{63.7} & 47/30 & 0.068 \\
Multi-constraint & 549 & 58.7 & 56.8 & \best{63.4} & 95/69 & 0.051 \\
\midrule
All & 824 & 59.2 & 57.7 & \best{64.4} & 137/94 & 0.006 \\
\bottomrule
\end{tabular}}
\end{table}

\section{Robustness of the Main Comparison}
\label{app:robust}

Table~\ref{tab:robust} repeats the comparison between the top-3 evidence and \method{} under four robustness checks. The longer output limit raises the reader's limit from 200 to 16,384 tokens, the PixelRAG setting. The updated-answer grader receives the Wikipedia page that the question was written from and also accepts an answer that this page states as the current one, which removes the penalty for references that have gone out of date since the datasets were built. The supported-reference subset keeps only the questions whose source page supports the reference answer. The held-out subset keeps the 4,851 questions outside the first 1,000 of each dataset, which our earlier pipeline experiments had used.

We also sampled 50 remaining \method{} failures per dataset and labeled their cause. An outdated or invalid reference or a grading error accounts for 72\%, 50\%, and 12\% of the \nq, \nqt, and \sqa samples. Among the pipeline failures, reader errors on delivered evidence outnumber missing evidence on every dataset (63 against 20 in total).

\begin{table}[h]
\caption{The \method{} gain under four robustness checks (pooled accuracy, \%; all significant with $p<10^{-12}$).}
\label{tab:robust}
\centering
\setlength{\tabcolsep}{2.5pt}
\begin{tabular}{@{}p{0.64\columnwidth}ccl@{}}
\toprule
\headrow \textbf{Condition} & \textbf{Top-3} & \textbf{\method{}} & \textbf{Gain} \\
\midrule
Default (text pipeline) & 70.70 & 75.74 & \gain{+5.03} \\
Output limit 16,384 instead of 200 tokens & 70.95 & 76.03 & \gain{+5.08} \\
Updated-answer grader, text (7,378) & 73.50 & 78.06 & \gain{+4.55} \\
Updated-answer grader, screenshot (7,378) & 71.70 & 77.46 & \gain{+5.76} \\
Source page supports reference (6,245) & 79.34 & 84.48 & \gain{+5.14} \\
Held-out questions, screenshot (4,851) & -- & -- & \gain{+5.50} \\
\bottomrule
\end{tabular}
\end{table}

\section{Unit Blocks in a Screenshot Context}
\label{app:pixel}

Table~\ref{tab:pixelalloc} varies how the unit block enters the screenshot pipeline at the reader cost of three tiles. The text and image forms of the block give the same accuracy ($p=0.89$), and a unit image re-packed to the token cost of the tile it replaces keeps the gain. Allocating two of the three tile budgets to units is the best configuration, and units without any tile still beat three tiles by 4.9 points. With Gemma 4 31B as the reader, the text form of the block is 1.9 points more accurate than its image form.

\begin{table}[h]
\caption{Screenshot pipeline at the three-tile reader cost (pooled accuracy, \%, Qwen3.5-4B reader).}
\label{tab:pixelalloc}
\centering
\setlength{\tabcolsep}{4pt}
\begin{tabular}{@{}lcl@{}}
\toprule
\headrow \textbf{Reader evidence} & \textbf{Pooled} & \textbf{vs.\ 3 tiles} \\
\midrule
Top-3 tiles & 69.58 & \\
Top-2 tiles + unit block as text & 75.18 & \gain{+5.59} \\
Top-2 tiles + unit block as an image & 75.23 & \gain{+5.64} \\
Top-2 tiles + token-matched unit image & 74.70 & \gain{+5.12} \\
\oursrow Top-1 tile + units in two tile budgets & \best{75.91} & \gain{+6.33} \\
Units in three tile budgets & 74.51 & \gain{+4.93} \\
\bottomrule
\end{tabular}
\end{table}

\section{Processing Cost}
\label{app:cost}

Table~\ref{tab:cost} compares the processing that each structure-aware method adds. \method{} never times out; HtmlRAG and RAPTOR add GPU passes over every page.

\input{tables/cost}

\ifarxiv
\section{Additional Examples}
\label{app:examples}

Figure~\ref{fig:examples} shows one more gain and one loss from the \sqa run. In the first case, the retrieved chunks discuss the Nobel Peace Prize, and the unit block delivers the list item that names the Ig Nobel winner under its year heading. In the second case, the third chunk that \method{} displaced contained the answer, and the unit block did not: a fixed split occasionally trades away a useful chunk. Across the 7,851 questions, \method{} gains 620 answers and loses 225.

\begin{figure}[h]
\centering
\begin{examplebox}{Gain: evidence in a list under a year heading}
\rdr{\textbf{Q:} Who won the 1991 Ig Nobel Prize for Peace?\hfill Reference: \textbf{Edward Teller}}\\[2pt]
\ctx{List of Ig Nobel Prize winners › 1991:}\\
Literature: Erich von Däniken, visionary raconteur and author of Chariots of the Gods?, ...\\
\hl{Peace: Edward Teller}, father of the hydrogen bomb and first champion of the Star Wars weapons system, "for his lifelong efforts to change the meaning of peace as we know it".\\[2pt]
\rdr{Top-3 chunks: ``... there is no mention of an Ig Nobel Prize for 1991 ...'' \xmark\\
\method{}: ``Edward Teller won the 1991 Ig Nobel Prize for Peace.'' \cmark}
\end{examplebox}
\vspace{-3pt}
\begin{examplebox}{Loss: the displaced chunk held the answer}
\rdr{\textbf{Q:} What object did Helly use to attempt suicide in Season 1, Episode 4 of Severance?\hfill Reference: \textbf{extension cord}}\\[2pt]
\ctx{Helly R. › Overview:}\\
... She makes several attempts to permanently leave the floor, including a suicide attempt. ...\\[2pt]
\rdr{Top-3 chunks: ``Helly used an extension cord ...'' \cmark\\
\method{}: ``... hanging herself in an elevator ... It does not specify a particular object ...'' \xmark}
\end{examplebox}
\caption{A gain and a loss of \method{} on \sqa{}.}
\label{fig:examples}
\end{figure}
\fi

%% file: tables/remedies.tex
\begin{table}[h]
\caption{Standard remedies on the hybrid text pipeline, with the oracle of Table~\ref{tab:oracle} for scale. $\Delta$: accuracy change in points on \nq / \nqt / \sqa, or pooled.}
\label{tab:remedies}
\centering
\setlength{\tabcolsep}{1pt}
\begin{tabular}{@{}p{0.7\columnwidth}r@{}}
\toprule
\headrow \textbf{Remedy} & \textbf{$\Delta$ accuracy} \\
\midrule
Recall-oriented extraction, full re-index & +0.5 / +0.8 / +0.4 \\
Neighbor chunks of the top pages, reranked & +0.4 / +1.5 / +0.2 \\
Page title prefixed to each chunk & +0.6 / $-$0.4 / +0.7 \\
Reader budget 3K $\rightarrow$ 12K tokens of chunks & +2.1 pooled \\
Screenshot tiles, 3 $\rightarrow$ 14 tiles (12K tokens) & +2.2 pooled \\
\midrule
Oracle evidence chunk (715 affected questions) & \gain{\textbf{+29.4}} \\
\bottomrule
\end{tabular}
\end{table}

%% file: tables/ablation.tex
\begin{table}[h]
\caption{Ablations on the hybrid text pipeline (accuracy, \%). Shaded rows are the defaults; panels (b) and (c) share one. $^\dagger$ differs significantly from the default. Panel (a) ran in an earlier software environment (top-3 chunks: 70.82 pooled).}
\label{tab:ablation}
\centering
\setlength{\tabcolsep}{2.1pt}
\begin{tabular}{@{}lcccc@{}}
\toprule
\headrow \textbf{Variant} & \textbf{\nq} & \textbf{NQ-Tab.} & \textbf{\sqa} & \textbf{Pooled} \\
\midrule
\grouprow{5}{(a) What the third slot contains}
Third chunk & 64.78$^\dagger$ & 48.41$^\dagger$ & 83.61$^\dagger$ & 70.82 \\
Chunks with title prefix & 65.38$^\dagger$ & 48.03$^\dagger$ & 84.31$^\dagger$ & 71.33 \\
Bare units, no context & 66.49$^\dagger$ & 48.31$^\dagger$ & 85.99$^\dagger$ & 72.58 \\
\oursrow Completed units & 68.17 & \best{51.69} & 89.85 & 75.44 \\
+ units as a 4th item & \best{68.56} & 51.40 & \best{90.08} & \best{75.67} \\
\grouprow{5}{(b) Row construction and packing}
DOM-order rows, flat & \best{68.38} & 51.31 & 89.73 & 75.43 \\
Faithful grid, flat & 67.99 & \best{52.72} & 90.11 & 75.62 \\
Grid, context grouped & 68.11 & 52.43 & 90.43 & \best{75.77} \\
\oursrow Grid, table grouped & 67.84 & 52.34 & \best{90.63} & 75.74 \\
\grouprow{5}{(c) Budget allocation and pool size}
Joint ranking & 67.63 & 53.00 & 89.32$^\dagger$ & 75.16$^\dagger$ \\
Two unit slots & 67.96 & 51.78 & 90.22 & 75.53 \\
Units only & 67.60 & 50.75 & 87.61$^\dagger$ & 74.09$^\dagger$ \\
Anchors from top 5 & 67.99 & 52.06 & 90.28 & 75.61 \\
Anchors from top 20 & 67.78 & 51.50 & 90.89 & 75.71 \\
Prefilter $N{=}50$ & 67.81 & 52.25 & 89.99$^\dagger$ & 75.43 \\
Prefilter $N{=}200$ & 68.35 & 52.43 & 90.48 & 75.90 \\
+ sibling rows & 67.81 & 53.00 & 90.48 & 75.75 \\
$-$ redundant units & 68.11 & 52.34 & 89.41$^\dagger$ & 75.32 \\
\bottomrule
\end{tabular}
\end{table}

%% file: tables/cost.tex
\begin{table}[H]
\caption{Cost of bringing page structure to the reader. Times are per page on one CPU core unless noted.}
\label{tab:cost}
\centering
\setlength{\tabcolsep}{3.5pt}
\begin{tabular}{@{}lll@{}}
\toprule
\headrow \textbf{Method} & \textbf{Page processing} & \textbf{Models added} \\
\midrule
\oursrow \method{} & 0.015 s & none \\
Unstructured & 0.39 s & none \\
Docling & 7.9 s (1.4\% time out) & none \\
HtmlRAG & 68 GPU-s / question & embedder, pruner \\
RAPTOR & 1.0 GPU-s (summaries) & LLM, embedder \\
MinerU-HTML & 5.2 GPU-s (extraction) & extractor LM \\
\bottomrule
\end{tabular}
\end{table}

%% file: references.bib
@inproceedings{guu2020realm,
  title = {{REALM}: Retrieval-Augmented Language Model Pre-Training},
  author = {Kelvin Guu and Kenton Lee and Zora Tung and Panupong Pasupat and Ming-Wei Chang},
  booktitle = {Proc. ICML},
  pages = {3929--3938},
  year = {2020},
}

@misc{hua2026semantics,
    title={Semantics Delivery Network: Rethinking Web Retrieval Infrastructure for LLM Agents},
    author={Peichun Hua and Yunming Xiao},
    year={2026},
    eprint={2609.22486},
    archiveprefix={arXiv},
    primaryClass={cs.NI},
}

@inproceedings{lewis2020retrieval,
  title = {Retrieval-Augmented Generation for Knowledge-Intensive {NLP} Tasks},
  author = {Patrick Lewis and Ethan Perez and Aleksandra Piktus and Fabio Petroni and Vladimir Karpukhin and Naman Goyal and Heinrich K{\"u}ttler and Mike Lewis and Wen-tau Yih and Tim Rockt{\"a}schel and Sebastian Riedel and Douwe Kiela},
  booktitle = {Proc. NeurIPS},
  pages = {9459--9474},
  year = {2020},
}

@inproceedings{karpukhin2020dense,
  title = {Dense Passage Retrieval for Open-Domain Question Answering},
  author = {Vladimir Karpukhin and Barlas Oguz and Sewon Min and Patrick Lewis and Ledell Wu and Sergey Edunov and Danqi Chen and Wen-tau Yih},
  booktitle = {Proc. EMNLP},
  pages = {6769--6781},
  year = {2020},
  doi = {10.18653/v1/2020.emnlp-main.550},
}

@inproceedings{izacard2021leveraging,
  title = {Leveraging Passage Retrieval with Generative Models for Open Domain Question Answering},
  author = {Gautier Izacard and Edouard Grave},
  booktitle = {Proc. EACL},
  pages = {874--880},
  year = {2021},
  doi = {10.18653/v1/2021.eacl-main.74},
}

@misc{wang2026pixelrag,
  title = {{PIXELRAG}: Web Screenshots Beat Text for Retrieval-Augmented Generation},
  author = {Yichuan Wang and Zhifei Li and Zirui Wang and Paul Teiletche and Lesheng Jin and Matei Zaharia and Joseph E. Gonzalez and Sewon Min},
  year = {2026},
  eprint = {2606.28344},
  archiveprefix = {arXiv},
  primaryclass = {cs.IR},
}

@inproceedings{ma2024dse,
  title = {Unifying Multimodal Retrieval via Document Screenshot Embedding},
  author = {Xueguang Ma and Sheng-Chieh Lin and Minghan Li and Wenhu Chen and Jimmy Lin},
  booktitle = {Proc. EMNLP},
  pages = {6492--6505},
  year = {2024},
  doi = {10.18653/v1/2024.emnlp-main.373},
}

@inproceedings{faysse2025colpali,
  title = {{ColPali}: Efficient Document Retrieval with Vision Language Models},
  author = {Manuel Faysse and Hugues Sibille and Tony Wu and Bilel Omrani and Gautier Viaud and C{\'e}line Hudelot and Pierre Colombo},
  booktitle = {Proc. ICLR},
  year = {2025},
}

@inproceedings{yu2025visrag,
  title = {{VisRAG}: Vision-Based Retrieval-Augmented Generation on Multi-Modality Documents},
  author = {Shi Yu and Chaoyue Tang and Bokai Xu and Junbo Cui and Junhao Ran and Yukun Yan and Zhenghao Liu and Shuo Wang and Xu Han and Zhiyuan Liu and Maosong Sun},
  booktitle = {Proc. ICLR},
  year = {2025},
}

@misc{cho2024m3docrag,
  title = {{M3DocRAG}: Multi-modal Retrieval is What You Need for Multi-page Multi-document Understanding},
  author = {Jaemin Cho and Debanjan Mahata and Ozan Irsoy and Yujie He and Mohit Bansal},
  year = {2024},
  eprint = {2411.04952},
  archiveprefix = {arXiv},
  primaryclass = {cs.CV},
}

@misc{li2026qwenvlretrieval,
  title = {{Qwen3-VL-Embedding} and {Qwen3-VL-Reranker}: A Unified Framework for Multimodal Retrieval and Ranking},
  author = {Mingxin Li and Yanzhao Zhang and Dingkun Long and Keqin Chen and Sibo Song and Shuai Bai and Zhibo Yang and Pengjun Xie and An Yang and Dayiheng Liu and Jingren Zhou and Junyang Lin},
  year = {2026},
  eprint = {2601.04720},
  archiveprefix = {arXiv},
  primaryclass = {cs.CL},
}

@misc{zhang2025qwen3,
  title = {{Qwen3 Embedding}: Advancing Text Embedding and Reranking Through Foundation Models},
  author = {Yanzhao Zhang and Mingxin Li and Dingkun Long and Xin Zhang and Huan Lin and Baosong Yang and Pengjun Xie and An Yang and Dayiheng Liu and Junyang Lin and Fei Huang and Jingren Zhou},
  year = {2025},
  eprint = {2506.05176},
  archiveprefix = {arXiv},
  primaryclass = {cs.CL},
}

@inproceedings{barbaresi2021trafilatura,
  title = {Trafilatura: A Web Scraping Library and Command-Line Tool for Text Discovery and Extraction},
  author = {Adrien Barbaresi},
  booktitle = {Proc. ACL-IJCNLP: System Demonstrations},
  pages = {122--131},
  year = {2021},
  doi = {10.18653/v1/2021.acl-demo.15},
}

@inproceedings{jin2025flashrag,
  title = {{FlashRAG}: A Modular Toolkit for Efficient Retrieval-Augmented Generation Research},
  author = {Jiajie Jin and Yutao Zhu and Zhicheng Dou and Guanting Dong and Xinyu Yang and Chenghao Zhang and Tong Zhao and Zhao Yang and Ji-Rong Wen},
  booktitle = {Companion Proc. WWW},
  pages = {737--740},
  year = {2025},
  doi = {10.1145/3701716.3715313},
}

@inproceedings{tan2025htmlrag,
  title = {{HtmlRAG}: {HTML} is Better Than Plain Text for Modeling Retrieved Knowledge in {RAG} Systems},
  author = {Jiejun Tan and Zhicheng Dou and Wen Wang and Mang Wang and Weipeng Chen and Ji-Rong Wen},
  booktitle = {Proc. WWW},
  year = {2025},
  doi = {10.1145/3696410.3714546},
}

@misc{wang2025readerlm,
  title = {{ReaderLM-v2}: Small Language Model for {HTML} to Markdown and {JSON}},
  author = {Feng Wang and Zesheng Shi and Bo Wang and Nan Wang and Han Xiao},
  year = {2025},
  eprint = {2503.01151},
  archiveprefix = {arXiv},
  primaryclass = {cs.CL},
}

@misc{liu2025dripper,
  title = {Dripper: Token-Efficient Main {HTML} Extraction with a Lightweight {LM}},
  author = {Mengjie Liu and Jiahui Peng and Pei Chu and Jiantao Qiu and Ren Ma and He Zhu and Rui Min and Lindong Lu and Wenchang Ning and Linfeng Hou and Kaiwen Liu and Yuan Qu and Zhenxiang Li and Chao Xu and Zhongying Tu and Wentao Zhang and Conghui He},
  year = {2025},
  eprint = {2511.23119},
  archiveprefix = {arXiv},
  primaryclass = {cs.CL},
}

@article{robertson2009bm25,
  title = {The Probabilistic Relevance Framework: {BM25} and Beyond},
  author = {Stephen Robertson and Hugo Zaragoza},
  journal = {Found. Trends Inf. Retr.},
  volume = {3},
  number = {4},
  pages = {333--389},
  year = {2009},
  doi = {10.1561/1500000019},
}

@inproceedings{cormack2009rrf,
  title = {Reciprocal Rank Fusion Outperforms Condorcet and Individual Rank Learning Methods},
  author = {Gordon V. Cormack and Charles L. A. Clarke and Stefan B{\"u}ttcher},
  booktitle = {Proc. SIGIR},
  pages = {758--759},
  year = {2009},
  doi = {10.1145/1571941.1572114},
}

@misc{chen2024bgem3,
  title = {{BGE M3-Embedding}: Multi-Lingual, Multi-Functionality, Multi-Granularity Text Embeddings Through Self-Knowledge Distillation},
  author = {Jianlv Chen and Shitao Xiao and Peitian Zhang and Kun Luo and Defu Lian and Zheng Liu},
  year = {2024},
  eprint = {2402.03216},
  archiveprefix = {arXiv},
  primaryclass = {cs.CL},
}

@article{johnson2021faiss,
  title = {Billion-Scale Similarity Search with {GPUs}},
  author = {Jeff Johnson and Matthijs Douze and Herv{\'e} J{\'e}gou},
  journal = {IEEE Trans. Big Data},
  volume = {7},
  number = {3},
  pages = {535--547},
  year = {2021},
  doi = {10.1109/TBDATA.2019.2921572},
}

@article{kwiatkowski2019natural,
  title = {Natural Questions: A Benchmark for Question Answering Research},
  author = {Tom Kwiatkowski and Jennimaria Palomaki and Olivia Redfield and Michael Collins and Ankur P. Parikh and Chris Alberti and Danielle Epstein and Illia Polosukhin and Jacob Devlin and Kenton Lee and Kristina Toutanova and Llion Jones and Matthew Kelcey and Ming-Wei Chang and Andrew M. Dai and Jakob Uszkoreit and Quoc Le and Slav Petrov},
  journal = {Trans. Assoc. Comput. Linguistics},
  volume = {7},
  pages = {452--466},
  year = {2019},
  doi = {10.1162/tacl_a_00276},
}

@inproceedings{herzig2021nqtables,
  title = {Open Domain Question Answering over Tables via Dense Retrieval},
  author = {Jonathan Herzig and Thomas M{\"u}ller and Syrine Krichene and Julian Eisenschlos},
  booktitle = {Proc. NAACL},
  pages = {512--519},
  year = {2021},
  doi = {10.18653/v1/2021.naacl-main.43},
}

@misc{wei2024simpleqa,
  title = {Measuring Short-Form Factuality in Large Language Models},
  author = {Jason Wei and Nguyen Karina and Hyung Won Chung and Yunxin Joy Jiao and Spencer Papay and Amelia Glaese and John Schulman and William Fedus},
  year = {2024},
  eprint = {2411.04368},
  archiveprefix = {arXiv},
  primaryclass = {cs.CL},
}

@misc{qwen2026qwen35,
  title = {{Qwen3.5}: Towards Native Multimodal Agents},
  author = {{Qwen Team}},
  year = {2026},
  howpublished = {Technical report},
  url = {https://qwen.ai/blog?id=qwen3.5}
}

@misc{openai-agent,
  author       = {{OpenAI}},
  title        = {Introducing {ChatGPT} Agent: Bridging Research and Action},
  howpublished = {OpenAI Blog},
  year         = {2025},
  url          = {https://openai.com/index/introducing-chatgpt-agent/}
}

@inproceedings{sarthi2024raptor,
  title = {{RAPTOR}: Recursive Abstractive Processing for Tree-Organized Retrieval},
  author = {Parth Sarthi and Salman Abdullah and Aditi Tuli and Shubh Khanna and Anna Goldie and Christopher D. Manning},
  booktitle = {Proc. ICLR},
  year = {2024}
}

@misc{gunther2024late,
  title = {Late Chunking: Contextual Chunk Embeddings Using Long-Context Embedding Models},
  author = {Michael G{\"u}nther and Isabelle Mohr and Daniel James Williams and Bo Wang and Han Xiao},
  year = {2024},
  eprint = {2409.04701},
  archiveprefix = {arXiv},
  primaryclass = {cs.CL}
}

@misc{sturua2024jina,
  title = {jina-embeddings-v3: Multilingual Embeddings With Task {LoRA}},
  author = {Saba Sturua and Isabelle Mohr and Mohammad Kalim Akram and Michael G{\"u}nther and Bo Wang and Markus Krimmel and Feng Wang and Georgios Mastrapas and Andreas Koukounas and Nan Wang and Han Xiao},
  year = {2024},
  eprint = {2409.10173},
  archiveprefix = {arXiv},
  primaryclass = {cs.CL}
}

@inproceedings{wang2023query2doc,
  title = {{Query2doc}: Query Expansion with Large Language Models},
  author = {Liang Wang and Nan Yang and Furu Wei},
  booktitle = {Proc. EMNLP},
  pages = {9414--9423},
  year = {2023},
}

@inproceedings{gao2023hyde,
  title = {Precise Zero-Shot Dense Retrieval without Relevance Labels},
  author = {Luyu Gao and Xueguang Ma and Jimmy Lin and Jamie Callan},
  booktitle = {Proc. ACL},
  pages = {1762--1777},
  year = {2023},
}

@inproceedings{abduljaleel2004rm3,
  title = {{UMass} at {TREC} 2004: Novelty and {HARD}},
  author = {Nasreen Abdul-Jaleel and James Allan and W. Bruce Croft and Fernando Diaz and Leah Larkey and Xiaoyan Li and Mark D. Smucker and Courtney Wade},
  booktitle = {Proc. TREC},
  year = {2004}
}

@misc{lassance2024spladev3,
  title = {{SPLADE-v3}: New Baselines for {SPLADE}},
  author = {Carlos Lassance and Herv{\'e} D{\'e}jean and Thibault Formal and St{\'e}phane Clinchant},
  year = {2024},
  eprint = {2403.06789},
  archiveprefix = {arXiv},
  primaryclass = {cs.IR}
}

@misc{jin2025searchr1,
  title = {{Search-R1}: Training {LLMs} to Reason and Leverage Search Engines with Reinforcement Learning},
  author = {Bowen Jin and Hansi Zeng and Zhenrui Yue and Jinsung Yoon and Sercan Arik and Dong Wang and Hamed Zamani and Jiawei Han},
  year = {2025},
  eprint = {2503.09516},
  archiveprefix = {arXiv},
  primaryclass = {cs.CL}
}

@misc{nakano2021webgpt,
  title = {{WebGPT}: Browser-Assisted Question-Answering with Human Feedback},
  author = {Reiichiro Nakano and Jacob Hilton and Suchir Balaji and Jeff Wu and Long Ouyang and Christina Kim and Christopher Hesse and Shantanu Jain and Vineet Kosaraju and William Saunders and others},
  year = {2021},
  eprint = {2112.09332},
  archiveprefix = {arXiv},
  primaryclass = {cs.CL},
}

@inproceedings{vu2023freshllms,
  title = {{FreshLLMs}: Refreshing Large Language Models with Search Engine Augmentation},
  author = {Tu Vu and Mohit Iyyer and Xuezhi Wang and Noah Constant and Jerry W. Wei and Jason Wei and Chris Tar and Yun-Hsuan Sung and Denny Zhou and Quoc V. Le and Thang Luong},
  booktitle = {Findings of ACL},
  pages = {13697--13720},
  year = {2024},
  doi = {10.18653/v1/2024.findings-acl.813}
}

@misc{song2025r1searcher,
  title = {{R1-Searcher++}: Incentivizing the Dynamic Knowledge Acquisition of {LLMs} via Reinforcement Learning},
  author = {Huatong Song and Jinhao Jiang and Wenqing Tian and Zhipeng Chen and Yuhuan Wu and Jiahao Zhao and Yingqian Min and Wayne Xin Zhao and Lei Fang and Ji-Rong Wen},
  year = {2025},
  eprint = {2505.17005},
  archiveprefix = {arXiv},
}

@misc{liu2026agenticr,
  title = {{Agentic-R}: Learning to Retrieve for Agentic Search},
  author = {Wenhan Liu and Xinyu Ma and Yutao Zhu and Yuchen Li and Daiting Shi and Dawei Yin and Zhicheng Dou},
  year = {2026},
  eprint = {2601.11888},
  archiveprefix = {arXiv},
}

@inproceedings{wu2025webwalker,
  title = {{WebWalker}: Benchmarking {LLMs} in Web Traversal},
  author = {Jialong Wu and Wenbiao Yin and Yong Jiang and Zhenglin Wang and Zekun Xi and Runnan Fang and Linhai Zhang and Yulan He and Deyu Zhou and Pengjun Xie and Fei Huang},
  booktitle = {Proc. ACL},
  pages = {10290--10305},
  year = {2025},
}

@inproceedings{krishna2025frames,
  title = {Fact, Fetch, and Reason: A Unified Evaluation of Retrieval-Augmented Generation},
  author = {Satyapriya Krishna and Kalpesh Krishna and Anhad Mohananey and Steven Schwarcz and Adam Stambler and Shyam Upadhyay and Manaal Faruqui},
  booktitle = {Proc. NAACL},
  year = {2025},
}

@article{trivedi2022musique,
  title = {{MuSiQue}: Multihop Questions via Single-hop Question Composition},
  author = {Harsh Trivedi and Niranjan Balasubramanian and Tushar Khot and Ashish Sabharwal},
  journal = {Trans. Assoc. Comput. Linguistics},
  volume = {10},
  pages = {539--554},
  year = {2022}
}

@misc{auer2024docling,
  title = {Docling Technical Report},
  author = {Christoph Auer and Maksym Lysak and Ahmed Nassar and Michele Dolfi and Nikolaos Livathinos and Panos Vagenas and Cesar Berrospi Ramis and Matteo Omenetti and Fabian Lindlbauer and Kasper Dinkla and others},
  year = {2024},
  eprint = {2408.09869},
  archiveprefix = {arXiv},
  primaryclass = {cs.CL}
}

@misc{unstructured,
  title = {Unstructured: Open-Source Pre-Processing Tools for Unstructured Data},
  author = {{Unstructured-IO}},
  year = {2024},
  url = {https://github.com/Unstructured-IO/unstructured}
}

@misc{liu2022llamaindex,
  title = {{LlamaIndex}},
  author = {Jerry Liu},
  year = {2022},
  url = {https://github.com/run-llama/llama_index}
}

@inproceedings{yang2017anserini,
  title = {Anserini: Enabling the Use of {Lucene} for Information Retrieval Research},
  author = {Peilin Yang and Hui Fang and Jimmy Lin},
  booktitle = {Proc. SIGIR},
  pages = {1253--1256},
  year = {2017}
}

@article{mcnemar1947,
  title = {Note on the Sampling Error of the Difference Between Correlated Proportions or Percentages},
  author = {Quinn McNemar},
  journal = {Psychometrika},
  volume = {12},
  number = {2},
  pages = {153--157},
  year = {1947}
}

@inproceedings{asai2024selfrag,
  title = {{Self-RAG}: Learning to Retrieve, Generate, and Critique through Self-Reflection},
  author = {Akari Asai and Zeqiu Wu and Yizhong Wang and Avirup Sil and Hannaneh Hajishirzi},
  booktitle = {Proc. ICLR},
  year = {2024}
}

@inproceedings{yao2023react,
  title = {{ReAct}: Synergizing Reasoning and Acting in Language Models},
  author = {Shunyu Yao and Jeffrey Zhao and Dian Yu and Nan Du and Izhak Shafran and Karthik Narasimhan and Yuan Cao},
  booktitle = {Proc. ICLR},
  year = {2023}
}

@inproceedings{chen2024densex,
  title = {Dense {X} Retrieval: What Retrieval Granularity Should We Use?},
  author = {Tong Chen and Hongwei Wang and Sihao Chen and Wenhao Yu and Kaixin Ma and Xinran Zhao and Hongming Zhang and Dong Yu},
  booktitle = {Proc. EMNLP},
  year = {2024},
}

@inproceedings{xu2024recomp,
  title = {{RECOMP}: Improving Retrieval-Augmented {LMs} with Context Compression and Selective Augmentation},
  author = {Fangyuan Xu and Weijia Shi and Eunsol Choi},
  booktitle = {Proc. ICLR},
  year = {2024}
}

@inproceedings{jiang2024longllmlingua,
  title = {{LongLLMLingua}: Accelerating and Enhancing {LLMs} in Long Context Scenarios via Prompt Compression},
  author = {Huiqiang Jiang and Qianhui Wu and Xufang Luo and Dongsheng Li and Chin-Yew Lin and Yuqing Yang and Lili Qiu},
  booktitle = {Proc. ACL},
  year = {2024},
}

@inproceedings{sui2024table,
  title = {Table Meets {LLM}: Can Large Language Models Understand Structured Table Data? A Benchmark and Empirical Study},
  author = {Yuan Sui and Mengyu Zhou and Mingjie Zhou and Shi Han and Dongmei Zhang},
  booktitle = {Proc. WSDM},
  pages = {645--654},
  year = {2024}
}

@misc{anthropic2024contextual,
  title = {Introducing Contextual Retrieval},
  author = {{Anthropic}},
  year = {2024},
  howpublished = {Anthropic Engineering Blog},
  url = {https://www.anthropic.com/engineering/contextual-retrieval}
}

@misc{google2026gemma4,
  title = {Gemma 4 Model Card},
  author = {{Google DeepMind}},
  year = {2026}
}

@misc{kiwix2025zim,
  title = {Wikipedia {ZIM} Archive \texttt{wikipedia\_en\_all\_maxi\_2025-08}},
  author = {{Kiwix}},
  year = {2025},
  url = {https://library.kiwix.org}
}

@misc{nogueira2019doctttttquery,
  title = {From doc2query to {docTTTTTquery}},
  author = {Rodrigo Nogueira and Jimmy Lin},
  year = {2019},
  howpublished = {Technical report, University of Waterloo}
}

@inproceedings{gospodinov2023doc2query,
  title = {{Doc2Query}-{}-: When Less is More},
  author = {Mitko Gospodinov and Sean MacAvaney and Craig Macdonald},
  booktitle = {Proc. ECIR},
  year = {2023}
}
